\documentclass[10pt]{paper}

\usepackage{amssymb}
\usepackage{amsmath}
\usepackage{epsfig}
\usepackage{graphics}
\usepackage{graphicx}
\usepackage{a4wide}

\newcommand{\be}{\begin{equation}}
\newcommand{\ee}{\end{equation}}
\newcommand{\bea}{\begin{eqnarray}}
\newcommand{\eea}{\end{eqnarray}}
\newcommand{\nn}{\nonumber}

\newcommand{\qe}{\varepsilon}
\newcommand{\qa}{\alpha}
\newcommand{\qb}{\beta}
\newcommand{\qg}{\gamma}
\newcommand{\qd}{\delta}
\newcommand{\qD}{\Delta}
\newcommand{\qh}{\eta}
\newcommand{\qj}{\psi}

\newcommand{\qk}{\kappa}
\newcommand{\ql}{\lambda}
\newcommand{\qr}{\rho}
\newcommand{\qs}{\sigma}
\newcommand{\qt}{\tau}
\newcommand{\qY}{\Theta}
\newcommand{\qf}{\varphi}

\newcommand{\qo}{\omega}
\newcommand{\qO}{\Omega}
\newcommand{\tri}{\triangle}
\newcommand{\Erf}{{\rm Erf}}

\newcommand{\naar}{\rightarrow}
\newcommand{\half}{\mbox{$\textstyle \frac{1}{2}$}}

\newcommand{\rd}{{\rm d}}
\newcommand{\prob}{{\mathbb P}}
\newcommand{\EE}{{\mathbb E}}

\newcommand{\EyXp}{{\mathbb E}_{yXp}}
\newcommand{\Ep}{{\mathbb E}_{p}}

\newcommand{\gnulT}{g_0^{\rm T}}
\newcommand{\goneT}{g_1^{\rm T}}
\newcommand{\fT}{f^{\rm T}}
\newcommand{\intp}{\int_{t}^{1-t}\!\!\!\!\!\!\!{\rm d}p\;}
\newcommand{\Erfcinv}{{\rm Erfc}^{\rm inv}}

\newtheorem{definition}{\rm\bf Definition}
\newtheorem{lemma}{\rm\bf Lemma}
\newtheorem{theorem}{\rm\bf Theorem}
\newtheorem{corollary}{\rm\bf Corollary}

\begin{document}

\title{Tardos fingerprinting is better than we thought}
\author{B. \v{S}kori\'{c}, T.U. Vladimirova,
        M. Celik, J.C. Talstra}

\maketitle

\begin{abstract}
Tardos has proposed a randomized fingerprinting code that is provably secure 
against collusion attacks. 
We revisit his scheme and show that 
it has significantly better performance than suggested in the original paper. 
First, we introduce variables in place of Tardos' hard-coded constants and 
we allow for an independent choice of the desired false positive and false negative error rates.
Following through Tardos' 
proofs with these modifications, we show that the code length can be reduced by more than 
a factor of two in typical content distribution applications where high false 
negative rates can be tolerated. 
Second, we study the statistical properties of the code.  
Under some reasonable assumptions, the accusation sums can be 
regarded as Gaussian-distributed stochastic variables.
In this approximation, the desired error rates are achieved by a code length twice shorter than in the first approach.
Overall, typical false positive and false negative error rates may be achieved 
with a code length approximately 5 times shorter than in the original construction.
\end{abstract}


\section{Introduction}

\subsection{Digital fingerprinting}
Digital content, such as songs, photographs or movies, can be copied inexpensively without any loss of quality. Moreover, these copies can be easily redistributed without permission from the original rights holders. According to \cite{RW2004}, 
unauthorized sharing of music on peer-to-peer (P2P) networks in college campuses reduces the potential revenue of the recoding industry by as much as 20\%.

One way of countering unauthorized redistribution is to uniquely mark each individual instance of the originally distributed content, so that the recipient (`user') can be identified if that content appears on a P2P network. 
The authorized distributor (or the content owner) embeds a unique mark, also called a `forensic watermark' or a `fingerprint', into each instance of the content before transmitting it to the user. The embedding algorithm ensures that the  mark is imperceptible, i.e. the quality of the content is not degraded by the mark. Moreover, the location and nature of the mark is kept secret from the user to prevent him from locating and altering the mark. 
Typically the mark consists of a set of symbols from a $q$-ary alphabet, where each symbol is embedded into a different part of the content, e.g. different scenes in a movie. When an unauthorized copy is found, the content owner, knowing all the details, can detect the mark and identify the source of the unauthorized copy. 
Forensic watermarking has already been successfully applied in practice~\cite{AP2004}.

\subsection{Collusion resistance}
A group of recipients (called `colluders' or `a coalition') can collaborate to escape identification.
Comparing their content copies, they can find the locations where their content, and thus their marks, differ. These locations are called the `detectable positions'.
By cleverly manipulating the content at those locations, the colluders can attempt to create a version of the content that cannot be traced back to any of them. Such an attack is called a {\em collusion attack}.

Collusion attacks and fingerprinting schemes that show resistance to these attacks have been studied 
since the late~1990's. The often used {\em marking condition} assumes that the colluders are unable to 
change the symbols (marks) in undetectable positions. Under the marking condition, 
one can distinguish between the following attack models, which differ in the type of manipulation 
the attackers are allowed to perform: 
\begin{itemize}
\item
The {\em restricted digit model} or {\em narrow-case model} allows the colluders only to 
`mix and match', i.e. to replace a symbol in a detectable position
by any of the symbols they have received in that position.
\item
The {\em unreadable digit model} allows for slightly stronger attacks. The attackers can also 
introduce an unreadable symbol `?' in detectable positions.
\item
The {\em arbitrary digit model} allows for even stronger attacks. The attackers can put any (arbitrary) $q$-ary symbol (but not the unreadable symbol `?') in the detectable positions.
\item
The {\em general digit model} allows the attackers to put any symbol, including the unreadable 
symbol `?', in the detectable positions.
\end{itemize}
In the case of a binary alphabet all four attack models are equivalent. The content owner can map the 
unreadable symbol `?' to either of the binary symbols without loss of generality.

Video fingerprinting applications face a number of severe constraints in practice.
First of all, there is a limit 
on the number of locations ($m$) suitable for watermark embedding.
A typical fingerprinting system can 
reliably extract approximately seven bits per minute of video content~\cite{DCI2007}.
Furthermore, constraints on decoding complexity and perceptual quality limit the number of different 
symbols that can be embedded in each location. Hence, the alphabet size $q$ for a fingerprint code 
is limited (typically $q\leq 16$). 
Finally, mass market content distribution systems need 
to accommodate a very large number of users (e.g. millions or even hundreds of millions).
Under these constraints, the authorized distributor is interested in the fingerprint code which can 
resist the largest coalition size~($c_0$).

In the last decade, various fingerprint codes have been proposed. Some of these codes are deterministic, i.e. 
they can identify at least one member of the coalition with certainty, without the danger of accusing an innocent user. For instance, Identifiable Parent Property (IPP) codes proposed in~\cite{HvLLT1998} are deterministic. However, the scheme is limited to a coalition size of two. 
In~\cite{SSW2001}, Staddon et al. proved the existence of a deterministic fingerprinting code which is resistant against $c_0$ colluders. The code is of length $m=c_0^2\log_q(n)$, where $n$ is the number of users.
However, it requires an impractically large alphabet size, $q\geq m-1$. 
Another deterministic scheme, presented in~\cite{CFNB2000}, has a similar length $m=4c_0^2\log n$ with a smaller alphabet size $q=2c_0^2$. Still, the alphabet size quickly becomes prohibitive for mass market 
content distribution systems. 

When the application can tolerate a nonzero probability of error, randomized fingerprinting 
codes with smaller, even binary, alphabets can be used. In a typical fingerprinting application, 
the most important type of error is the False Positive (FP) error, where an innocent user gets accused. 
The probability of such an event must be extremely small; otherwise all accusations become dubious, making 
the whole fingerprinting scheme unworkable.
We will denote by $\qe_1$ the probability that a specific innocent user gets accused.
The notation $\qh$ is used for the probability that there exist innocent users among the accused.
The second type of error is the False Negative (FN) error, where the scheme fails to accuse any of the colluders. In practical content distribution applications, fairly large FN error probabilities 
can be tolerated, as the content owner can collect evidence from multiple pieces of content over a period of time.  
We denote the FN error probability by $\qe_2$.

In \cite{BS1998}, Boneh and Shaw presented a binary ($q=2$) randomized code with length 
$m={\cal O}(c_0^4 \ln \frac{n}{\qh}\ln\frac{1}{\qh})$, which uses concatenation of a 
partly randomized inner code with an outer code. 
They also proved a lower bound on the required length for any binary code 
that is resistant against $c_0$ colluders: 
$m\geq\half(c_0-3)\ln\frac{1}{c_0\qh}$.
In \cite{PSS2003}, Peikert et al. proved a tighter lower bound for a restricted class of 
codes with a limited number of `column types':
$m=\qO(c_0^2\ln\frac{1}{c_0\qh})$.

In \cite{Tardos}, Tardos further tightened the lower bound for the arbitrary digit model and the unreadable digit model: 
$m=\qO(c_0^2\ln\frac{1}{\qe_1})$ for arbitrary alphabets.
In the same paper, he described a fully randomized binary fingerprinting code achieving this lower bound. This code has length $m=100c_0^2\lceil\ln\frac{1}{\qe_1}\rceil$. 

While \cite{Tardos} proposed a practical code with optimal behavior for large $c_0$, it leaves a number of open questions:
\begin{itemize}
\item
The author uses a number of arbitrary-looking constants in his proofs, such as $20$, $\frac{1}{10c_0}$ $\frac{1}{20c_0}$, $\frac{1}{300c_0}$. Similarly, the constant 100 appears in the minimum code length expression $m=100c_0^2\lceil\ln\frac{1}{\qe_1}\rceil$. While these numbers allow for important 
properties to be proven, it is not at all clear if they have been chosen in an optimal way.
\item
He also makes seemingly arbitrary choices for the accusation weight function---which specifies how strongly to accuse a user per symbol if his symbol is equal to the symbol found in a pirated copy---and the distribution function---which specifies probabilities used in generating the random code words,---hinting that they are optimal, but not providing a proof. 
\item
In the proofs, the FN error probability $\qe_2$ is coupled to the FP probability~$\qe_1$. While Tardos 
remarks that they can be decoupled, it is not clear how each exactly influences the code length on its own.
Furthermore, the coupling is such that FN rate is much smaller than the FP rate, $\qe_2\ll\qe_1$. 
As mentioned earlier, in practical applications the opposite $\qe_2\gg \qe_1$  may be desirable. This
opposite case is not studied. 
\end{itemize}


\subsection{Contributions and outline}
In this paper, we provide answers to the aforementioned issues with Tardos' construction, 
which were left open in~\cite{Tardos}.

\begin{itemize}
\item
In Section~\ref{secnotation}, we generalize the Tardos' construction, 
introducing variables in place of numerical constants and generic functions instead 
of the functions specified in~\cite{Tardos}. We state `Soundness' and `Completeness' properties, which 
specify the desired FP and FN error conditions.
Our approach is similar in spirit to Hagiwara et al.'s in \cite{HHI2006}, but our results are not restricted to small coalitions. 
\item
In Section~\ref{secmainresult}, we state the conditions on the construction parameters for a scheme that satisfies both the `Soundness' and `Completeness' properties. These conditions are derived in  Sections~\ref{secProp1} and~\ref{secProp2}, respectively. We employ a proof method very similar to~\cite{Tardos}, but we specifically do not couple FP and FN error rates.
In Section~\ref{secboundnu}, the results of the preceding subsections are combined to arrive at a 
condition for the code length.
\item
In Section \ref{secgoptimal}, we show that Tardos' choice for the accusation weight function is optimal.
In Section~\ref{secfoptimal}, we show that his choice for the distribution function used in generating the random code words is optimal within a limited class of functions.
\item
In Section \ref{sec4pi2}, we arrive at the smallest possible code length parameter value, specific for our proof method, that allows for Soundness and Completeness.
In the case of large $c_0$,
this value lies slightly above $4\pi^2$, which is a significant improvement
over the original value~`100'.
\item
In Section \ref{secnumerics}, we present the results of a numerical search for model parameters 
that yield the shortest possible code length. For certain realistic choices of $n$, $c_0$, 
$\qe_1$ and $\qe_2$, we find that the code length can be reduced by a factor of two or more.
For large $\qe_2$ our theoretical large-$c_0$ result seems to be approached.
\item
In Section \ref{secnumerics}, we further see that the code length depends only weakly on $\qe_2$, 
as was also mentioned by Tardos. Nonetheless, we observe a significant advantage for decoupling the FP and FN 
error probabilities. When Tardos' coupling of $\qe_2=\qe_1^{c_0/4}$ is enforced, the code length constant `100' can be reduced only to approximately~90. When the rates are decoupled and the FN rate is allowed to increase, for fixed $\qe_1$ and $c_0$ values,
the constant can be reduced to approximately~45.
\end{itemize}
We also study the statistical properties of the scheme. The results give us an insight into the `average' 
behavior of the scheme and are much simpler than the expressions involved in the formal proofs.
\begin{itemize}
\item
In Section \ref{statMainResult}, by modeling the accusation sums as normally distributed stochastic variables (an approximation motivated by the 
Central Limit Theorem), we state a simple approximate condition for the code length based on 
the coalition size, FP and FN error rates.
\item 
In Sections \ref{secstatSj} and \ref{secstatS}, we 
compute the mean and the variance of the accusation sums for innocent users and the coalition, without 
any assumptions on their distribution. Similarly, we derive conditions on the code 
length and the accusation threshold based on the desired false positive and false negative 
error rates, in Section~\ref{seccdep}.
\item
In Section~\ref{secstrategy}, we identify an `extremal' colluder strategy.
It maximizes our expression for the minimally required code length.
The strategy is to output a `1' whenever this is allowed by the marking condition.
\item
In Section~\ref{secfinalgaussian}, we assume that the probability distributions of the
accusation sums are perfectly Gaussian. Using this approximation, 
we show that (for large $c_0$) error rates $\qe_1$, $\qe_2$ can be achieved with code length 
$m\approx 2\pi^2 c_0^2\ln\frac{1}{\qe_1}$.
The Gaussian approximation is justified in Appendix~\ref{appCLT}.
\end{itemize}


\section{Tardos revisited}
\label{secrevisit}
\subsection{Generalization}
\label{secgeneralisation}

In~\cite{Tardos}, Tardos proposed a randomized fingerprinting code resilient against collusion attacks. 
His fingerprinting scheme is particularly known for its short code length. Nonetheless, due to a number 
of implicit parameter choices in~\cite{Tardos}, it is not clear if his explicit construction achieves the 
shortest possible code length allowed by the scheme. 
In this section, we generalize Tardos' construction in anticipation of 
our study in the following section. In particular, we make three generalizations. 
First, we replace 
various fixed numerical parameters by variables and investigate the conditions on these 
variables that allows us to still carry out the security proofs. 
This helps us 
to modify the system parameters so as to obtain even shorter code lengths. Second, we allow for the desired 
false positive and false negative error probabilities to be chosen independently. As opposed to the original 
construction where these probabilities were coupled, this generalization allows us to better align the scheme 
to practical content fingerprinting requirements. Third, we do not assume any specific form for the functions $f$, $g_1$ and $g_0$ (see Section~\ref{secnotation}) in Tardos' construction. The introduction of arbitrary functions $f$, $g_1$, $g_0$ leads to a lot of extra effort in carrying out the proofs. However, with this generalization, we can show that Tardos' choices for $g_1$, $g_0$ are 
optimal (for the proof technique employed in \cite{Tardos}), 
and that his choice for $f$ is likewise optimal within a specific class of smooth functions.

\subsection{Notation and definitions}
\label{secnotation}

We describe our generalized version of the binary Tardos fingerprinting scheme below. We adhere to the notation in \cite{Tardos} whenever  possible. Moreover, we denote the explicit parameter choices made by Tardos by the superscript~`T' to avoid confusion.

The fingerprinting scheme has~$n$ recipients (`users'). Each user is assigned a codeword of length~$m$. The 
set of colluders (the coalition) is denoted as~$C$, whereas the number of colluders is denoted as~$c$. 
The coalition size up to which the scheme has to be resistant is denoted as~$c_0$.
The content owner generates an $n\times m$ matrix $X$; the $j$-th row of $X$ is the codeword embedded in the content of user~$j$.
The part of $X$ received by the coalition is denoted as~$X_C$.
The colluders use a `$C$-strategy' $\qr$ to produce an  
$m$-bit string~$y=\qr(X_C)$ which ends up in the unauthorized copy. 
The strategy $\qr$ can be deterministic or stochastic. 
The content owner uses an accusation algorithm $\qs$. The output of the algorithm is a list of accused users.

The matrix $X$ is constructed in two phases. 
In the first phase a list of random numbers $p=\{p_i\}_{i=1}^m$ is generated, where $p_i\in [t,1-t]$, with $t$ a small parameter satisfying $c_0 t\ll 1$. The $p_i$ are independent and identically distributed according to a probability distribution function~$f$. The function $f$ is symmetric around $p_i=\half$ and heavily biased towards values of $p_i$ close to 0 and 1.
In the second phase, the columns of $X$ are filled by independently drawing random numbers $X_{ji}\in\{0,1\}$ with $\prob[X_{ji}=1]=p_i$.

Having spotted a copy with embedded mark $y$,
the content owner computes an `accusation sum' $S_j$ for each user $j$ according to 
\bea
		 S_j=\sum_{i=1}^m y_i U(X_{ji},p_i)
		 \hskip2mm &;& \hskip2mm 
		 U(X_{ji},p_i)=\left\{\begin{matrix}
		 g_1(p_i) \;\;{\rm if}\;\; X_{ji}=1
		  \cr
		 g_0(p_i) \;\;{\rm if}\;\; X_{ji}=0
		  \end{matrix}
		 \right.
\label{accuse}
\eea
where $g_0$ and $g_1$ are the `accusation functions', 
and $y_i$ denotes the $i$'th bit of~$y$.
The decision whether to accuse a user is taken as follows:
if $S_j>Z$, then accuse user~$j$. Hence
\be
	\qs(X,y,p)=\{j\; | \; S_j>Z\}.
\label{threshold}
\ee
We note some properties of this construction. If the received mark $y_i$ is zero, the accusation of 
user $j$ due to column~$i$ is neutral. If $y_i=1$, the accusation sum is updated by $g_{X_{ji}}(p_i)$, 
which is a measure of how much suspicion arises from observing $y_i$ for a given $X_{ji}$ and~$p_i$. 
The function $g_1(p)$ is positive and monotonically decreasing. The fact that user $j$ has received 
a `1' in that position, i.e. $X_{ji}=1$, adds to the suspicion. Moreover, the amount of suspicion decreases with increasing~$p_i$, as the symbol becomes more probable. 
The function $g_0(p)$ is negative and monotonically decreasing. Therefore, the fact that $X_{ji}=0$ detracts from the suspicion, and this becomes more pronounced for large $p_i$.
We impose two properties on the accusation functions $g_0$ and $g_1$, which become handy during our proofs. 
First, we want to have the expectation value of the accusation in each column to be zero. Therefore, the functions should satisfy $p g_1(p) + (1-p) g_0(p) = 0$. Furthermore, we want the weights of the accusation for $X=0$ and $X=1$ to be symmetric, since the function $f(p)$ also has $0\leftrightarrow 1$ symmetry. 
This is achieved by setting $g_0(p)=-g_1(1-p)$.
These two properties together imply that (a) $g_0$ can be computed from $g_1$ according to $g_0(p)=-g_1(p)\cdot p/(1-p)$ and
(b) $g_1$ on the interval $(\half,1-t)$ can be derived from $g_1$ on the interval $(t,\half)$ according to $g_1(p)=g_1(1-p)\cdot(1-p)/p$.
Hence it is necessary only to specify $g_1(p)$ for $p\in(t,\half)$.
\footnote{The constraint $g_1'(p)<0$ on the whole interval $(t,1-t)$ gives the additional condition $g_1'(p)>-\frac{g_1(p)}{p(1-p)}$.
This can be satisfied by writing
$g_1(p)=\exp \int^p \rd p'\; Q(p')$ with $-\frac{1}{p(1-p)}<Q(p)<0$.
}

The FP parameter $\qe_1$, chosen by the content owner, denotes the desired bound on the probability of having $j\in\qs(X,y,p)$ when a fixed user $j$ is innocent.
The FN parameter $\qe_2$, also chosen by the content owner, denotes the desired bound on the probability that $\qs(X,y,p)$ does not contain any guilty user.

The Tardos fingerprinting scheme uses a code length $m$ and a threshold $Z$ with the following scaling behavior\footnote{
Note that (\ref{defmZ}) has no explicit $\qe_2$-dependence.
The dependence enters implicitly through $A$ and $B$. However, as will be shown in Section~\ref{secnumerics}, 
the $\qe_2$-dependence vanishes in the limit of large~$c_0$. 
}
 as a function of $n$, $\qe_1$ and $c_0$:
\bea
		 m = A c_0^2 \left\lceil \ln 1/\qe_1\right\rceil
		 \hskip2mm &;& \hskip2mm 
		 Z = B c_0 \left\lceil \ln 1/\qe_1\right\rceil.
\label{defmZ}
\eea
Here we have introduced the parameters $A$ and~$B$, replacing Tardos' constants 100 and 20, respectively.
We want the scheme to satisfy the following properties.
\begin{definition}
\label{def:soundness}
`Soundness'\newline
Let $\qe_1\in(0,1)$ be a fixed constant and let $j$ be an arbitrary innocent user.
We say that the above described fingerprinting scheme is $\qe_1$-sound
if, for all coalitions $C\subseteq [n]\backslash\{j\}$, and for all $C$-strategies $\qr$,  
\be
	\prob[{\rm False\; positive}]=
	\prob[j\in\qs]< \qe_1.
\ee
\end{definition}
\begin{definition}
\label{def:completeness}
`Completeness' \newline
Let $\qe_2\in(0,1)$ and $c_0\in{\mathbb N}^+$ be fixed constants.
We say that the fingerprinting scheme is $(c_0,\qe_2)$-complete if,
for all coalitions $C$ of size $c\leq c_0$, and all $C$-strategies $\qr$,
\be
 		 \prob[{\rm False\; negative}]=
 		 \prob[C\cap \qs=\emptyset]<\qe_2.
\ee
\end{definition}
Tardos proved (for $c_0\geq 7$) that his scheme is Sound and Complete for the following very 
specific choice of parameters:\footnote{
In \cite{Tardos} the distribution $f$ was given in terms of a uniform random variable $r$, defined according to $p=\sin^2 r$.
}
\bea
	\fT(p)=\frac{1}{\pi-4t'}\frac{1}{\sqrt{p(1-p)}};
		 &\hskip3mm
		 \goneT(p)=\sqrt{\frac{1-p}{p}};
		 &\hskip3mm
		 \gnulT(p)=-\sqrt{\frac{p}{1-p}}
		 \nn\\
		 A^{\rm T}=100;
		 &
		 B^{\rm T}=20;
		 &
		 t^{\rm T}=\frac{1}{300c_0}
		 \nn\\
		 \qe_2^{\rm T}=\qe_1^{c_0/4} &&
\eea
where $t'=\arcsin\sqrt{t}$.


\section{Proving a shorter code length through parametrization}
\label{secreduceA}

\subsection{Main result}
\label{secmainresult}
The main aim of this paper is to
show that it is possible to satisfy Soundness and Completeness 
in Tardos' scheme
also with different choices, especially with a smaller parameter $A$ and hence shorter code length~$m$.
While the choice $A^{\rm T}=100$ does the job for $c_0\geq 7$, 
we will show that it can be reduced to a number slightly larger than $4\pi^2$ 
in the limit of large~$c_0$, when $\qe_2$ is {\em not} coupled to~$\qe_1$. 
Our results can be summarized in the form of the following theorem.
\begin{theorem}
\label{theorem:provablebound}
Let $\qe_1,\qe_2\in(0,1)$ be fixed parameters.
Let the cutoff parameter $t$ be parametrized as $t=\qt/c_0$, with $0<\qt\ll1$. 
Let $c_0$ satisfy
\be
	c_0\geq \frac{1}{\qt(3.4\pi)^2}.
\label{mainresultc0}
\ee
Let $\qo\ll 1$ be a positive constant.
Let the quantities $D$, $\qd$ and $\xi$ be defined as
\bea
\label{defD}
	D & := & e (\qo/1.7)^2 \nn\\
\label{defqd}
	\qd & := & 2\qt+\pi\qo+e^{1.7}\frac{\pi c_0}{\qo(1-D)}
	D^{\qt+\frac{1.7\sqrt{\qt}}{\qo}\sqrt{c_0-\qt}}
	\nn\\
\label{defxi}
	\xi&:=& \sqrt{1+\frac{1-\qd}{\pi\qo c_0}
	\cdot\frac{\ln\qe_2}{\ln\qe_1}}-1.
\eea
Let the length $m$ and the threshold $Z$ in the generalized Tardos scheme be parametrized according to (\ref{defmZ}).
Then it is possible to find functions $f(p)$, $g_1(p)$ such that the parameter setting
\bea
	A=4\pi^2\frac{(1+\xi)^2}{(1-\qd)^2} &;& 
	B=4\pi\frac{1+\xi}{1-\qd}.
\label{ABxiqd}
\eea
achieves $\qe_1$-soundness and $(c_0,\qe_2)$-completeness.
\end{theorem}

\begin{corollary}
\label{corol:mainresult}
In the limit of large $c_0$, $\qe_1$-soundness and $(c_0,\qe_2)$-completeness can be achieved by a code of length
\be
	m=4\pi^2\frac{1}{(1-2\qt-\pi\qo)^2}[1+{\cal O}(\frac{1}{c_0})]
	\cdot c_0^2 \ln\frac{1}{\qe_1}.
\ee
\end{corollary}

The proof of Theorem~\ref{theorem:provablebound} is given in the coming sections; we follow Tardos' proof method wherever possible. 
We derive the conditions for Soundness and Completeness in Sections~\ref{secProp1} and~\ref{secProp2}, respectively.
We combine these conditions in Section~\ref{secboundnu} and obtain 
the lowest possible value of $A$ that allows for Soundness and Completeness, depending
on the specific choice of the functions $f$ and~$g_1$.
In Section~\ref{secgoptimal}, we prove that $g_1=g_1^{\rm T}$ is the `optimal' choice, independent of~$f$, in the sense that this choice minimizes this lowest value of~$A$ (given the proof method). 
In Section~\ref{secfoptimal}, we argue that, given $g_1=g_1^{\rm T}$,
the choice $f=f^{\rm T}$ is `optimal' (in the same sense) within a limited class of functions.
Finally, in Section~\ref{sec4pi2}, we set $f=f^{\rm T}$ and $g_1=g_1^{\rm T}$ and complete the last step of the proof.


\subsection{Condition for Soundness}
\label{secProp1}

This section follows the lines of Tardos' proof of Theorem~1 in~\cite{Tardos}.
We derive an inequality for the scheme parameters from the requirement that the Soundness property holds.

First, an auxiliary variable $\qa_1>0$ is introduced for the purpose of applying the Markov inequality.
(Tardos uses $\qa_1^{\rm T}=1/(10c_0)$.) 
\be
	\prob[j\in\qs]=	 
	\prob[S_j>Z] = \prob[e^{\qa_1 S_j}>e^{\qa_1 Z}]
	< \frac{\EyXp[e^{\qa_1 S_j}]}{e^{\qa_1 Z}}.
\label{P1s1}
\ee
Here the notation $\EyXp$ denotes the expectation value 
computed by averaging over all 
stochastic degrees of freedom: 
the (possibly stochastic) $y_i$,
all the entries in~$X$
and the parameters $p_i$.
The expectation value in (\ref{P1s1}) is bounded by using the inequality\footnote{
One may ask why this crude inequality is used when our purpose is to squeeze the scheme's parameters as tightly as possible.
The answer is that
we do not want to deviate from \cite{Tardos} too much in this paper, as it would lead to even more bookkeeping than is already the case.
It would be interesting to determine the consequences of using an inequality of the more general form $e^u<1+ru+su^2$, with $r\geq 1$.
} 
$e^u\leq 1+u+u^2$, which holds for $u<1.7$.
Using the notation $u_i=g_{X_{ji}}(p_i)$, we can write
\be
	\EE_{X_j}[e^{\qa_1 S_j}]=
	\prod_{i: y_i=1}\EE_{X_{ji}}[ e^{\qa_1 u_i}]
	\leq \prod_{i: y_i=1}
	\left(  1+\qa_1\EE_{X_{ji}}[u_i]
	+\qa_1^2 \EE_{X_{ji}}[u_i^2] \right)
\label{P1s2}
\ee
Here $\EE_{X_j}$ stands for the expectation value 
over the $j$'th row of $X$, keeping the rest of $X$ fixed, and keeping 
$\{y_i\}$ and
$\{p_i\}$ fixed. Note that $y$ is independent of $X_j$ since the user $j$ is innocent.
The inequality (\ref{P1s2}) holds
as long as $\qa_1$ is so small that $\qa_1 u_i<1.7$ for all $i$ for which $y_i=1$.
This is automatically true for those columns where $X_{ji}=0$, since $g_0$ is negative. In the other columns, we need $\qa_1 g_1(p_i)<1.7$.
Since $p_i\geq t$ and $g_1$ is monotonously decreasing, we can satisfy the inequality for all $X$ by setting $\qa_1 <\qa_1^{\rm max}:= 1.7/g_1(t)$.

Next we further bound (\ref{P1s2}). Due to the property $pg_1(p)+(1-p)g_0(p)=0$ we have $\EE_{X_{ji}}[u_i]=0$. Thus we can write
\be
	\EE_{X_j}[e^{\qa_1 S_j}]\leq 
	\prod_{i: y_i=1}\left(  1 +\qa_1^2 \EE_{X_{ji}}[u_i^2] \right)
	\leq \prod_{i=1}^m\left(  1 +\qa_1^2 \EE_{X_{ji}}[u_i^2] \right).
\label{P1sinter}
\ee
Note that all dependence on the coalition strategy has disappeared in the last
expression.

Next we take the expectation value of (\ref{P1sinter}), for fixed $\{p_i\}$, over the remaining degrees of freedom in $X$ (all rows except $j$) and over $\{y_i\}_{i=1}^m$.
This has no effect on the last expression in (\ref{P1sinter}).
Finally we take the expectation value $\Ep$ w.r.t. the $p_i$ degrees of freedom.
We remind the reader that this amounts to multiplying with the distribution function $\prod_{i=1}^m f(p_i)$ and integrating over all $p_i$.
For ease of notation, we introduce the functional $\nu$, defined as
\be
	\nu:=\Ep\left[\EE_{X_{ji}}[u_i^2]\right]
	=\intp f(p)\left\{p[g_1(p)]^2+(1-p)[g_0(p)]^2
	\vphantom{\int}\right\}
	=2\int_t^{1/2}\!\!\!\!{\rm d}p\;
	f(p)\frac{p}{1-p}[g_1(p)]^2.
\label{defnu}
\ee
(With Tardos' choice for $g_1$, this evaluates to $\nu=\int_t^{1-t}{\rm d}pf(p)=1$).
We can then write
\be
	\EyXp[e^{\qa_1 S_j}]\leq \prod_{i=1}^m(1+\nu\qa_1^2)
	\leq e^{m\nu\qa_1^2}.
\label{P1s3}
\ee
In the last step we have used $1+u \leq e^u$, which holds for all $u$.
Substitution of (\ref{P1s3}) into (\ref{P1s1}) finally gives
\be
		 \prob[S_j>Z]< \exp(m\nu\qa_1^2-\qa_1 Z).
\label{P1res}
\ee
As (\ref{P1res}) holds for all $\qa_1$ in the allowed range, we can write
\be
	\prob[S_j>Z]< \min_{\qa_1\in(0,\qa_1^{\rm max})}
	\exp(m\nu\qa_1^2-\qa_1 Z).
\label{P1resminim}
\ee
The minimum of the parabola in the exponent lies at $\qa_1^*:=Z/(2m\nu)$.
Hence, the minimum in (\ref{P1resminim})
is obtained by setting $\qa_1=\qa_1^*$. Note that this is allowed only if $\qa_1^{\rm max}\geq\qa_1^*$; this condition can be rewritten as
\be
	c_0\geq \frac{Bg_1(t)}{3.4 \nu A}.
\label{ccondition1}
\ee
If $c_0$ is large enough for the condition to be satisfied,
then it holds that
\be
	\prob[S_j>Z] < \exp(m\nu[\qa_1^*]^2-\qa_1^* Z) 
	< \qe_1^{B^2/(4\nu A)}.
\label{P1resBA}
\ee
In the last inequality we have used $\qa_1^*=Z/(2m\nu)$ and the parametrization (\ref{defmZ}).
From (\ref{P1resBA}) and Definition~\ref{def:soundness} we conclude that, for $c_0$ large enough so that (\ref{ccondition1}) holds,
$\qe_1$-soundness can only be obtained if
\be
	A\leq \frac{B^2}{4\nu}.
\label{conditionright}
\ee


\subsection{Condition for Completeness}
\label{secProp2}
This section closely follows the proof of Theorem~2 in~\cite{Tardos}.
We derive an inequality for the scheme parameters from the Completeness requirement.
First, the coalition's accusation sum $S$ is defined,
\be
	S=\sum_{j\in C} S_j = \sum_{i=1}^m y_i
	\{ x_i g_1(p_i)+[c-x_i]g_0(p_i)  \}.
\label{defS}
\ee
Here $x_i=\sum_{j\in C} X_{ji}$ denotes the number of colluders that have a `1' at the $i$-th position of their codeword.
(The size of the coalition is $c=|C|$. We consider the case $c\leq c_0$.)
Since $S>cZ$ would imply that at least one colluder gets accused, the false negative error probability can be bounded by
\be
	\prob[C\cap \qs=\emptyset] \leq \prob[S\leq cZ].
\ee
An auxiliary constant $\qa_2>0$ is introduced for the purpose of using the Markov inequality 
(Tardos chooses $\qa_2^{\rm T}=1/(20c_0)$),
\be
	\prob[S\leq cZ] = \prob[e^{-\qa_2 S} \geq e^{-\qa_2 cZ}]
	\leq \frac{\EyXp[e^{-\qa_2 S}]}{e^{-\qa_2 cZ}}.
\ee
Upper bounding the expectation value $\EyXp[e^{-\qa_2 S}]$
is an arduous job.
The derivation is given in Appendix~\ref{appTheorem2}.
It turns out that $\EyXp[e^{-\qa_2 S}]<\exp(-\qa_2 m/L)$, with
$L>0$ a numerical constant, provided that $c$, $L$, $t$ and $\qa_2$ satisfy a complicated
condition (\ref{condition}).
(Tardos has $L^{\rm T}=4$).
Thus we can bound the false negative probability as
\be
	\prob[C\cap \qs=\emptyset]< \exp(\qa_2 cZ-\qa_2 m/L)
	\leq \exp(\qa_2 c_0 Z-\qa_2 m/L).
\label{P2res}
\ee
Substituting the parametrization (\ref{defmZ}) into (\ref{P2res}),
and restricting ourselves to the regime $A>LB$, we get
\be
	\prob[C\cap \qs=\emptyset]<\qe_1^{\qa_2 c_0^2[A/L-B]}.
\ee
Next we demand Completeness, $\prob[C\cap \qs=\emptyset]<\qe_2$.
This yields the following inequality,
\be
	A\geq LB+\frac{L}{\qa_2 c_0^2}\frac{\ln\qe_2}{\ln\qe_1}.
\label{conditionleft}
\ee 
Note that (\ref{conditionleft}) is valid only if the complicated condition (\ref{condition}) is satisfied.


\subsection{Conditions on the code length}
\label{secboundnu}
By combining the results of Sections~\ref{secProp1} and \ref{secProp2},
we derive a condition on the length parameter $A$ that is sufficient
for proving Soundness and Completeness using the current proof technique.
\begin{lemma}
\label{lemma:AnuL}
Let $\qe_1, \qe_2\in(0,1)$ be fixed parameters.
Let the code length $m$ and the threshold $Z$ in the generalized Tardos scheme be parametrized 
in terms of the $A$ and $B$ parameters
according to (\ref{defmZ}).
Let the threshold $t$ be set as $t=\qt/c_0$, with $\qt\ll 1$ a positive constant. 
Let $c_0$ be a fixed integer, satisfying
\be
	c_0 \geq \frac{1}{\qt(3.4 \nu\pi)^2}.
\label{constructLc0}
\ee
Let $\qa_2$ be the auxiliary parameter introduced in Section~\ref{secProp2}.
Let $L$ be the parameter introduced in Section~\ref{secProp2}, chosen such that the 
condition (\ref{condition}) can be satisfied by some value~$\qa_2$.
Let $\nu$ be the functional of $f$ and $g_1$ as defined by (\ref{defnu}).
Let $\qj$ be defined as
\be
	\qj=\sqrt{1+\frac{1}{\nu L\qa_2 c_0^2}\frac{\ln\qe_2}{\ln\qe_1}}-1.
\label{defqj}
\ee
Then the generalized Tardos scheme with
\bea
	A=4\nu L^2 (1+\qj)^2 &;& B=4\nu L(1+\qj)
\label{constructLAB}
\eea
is $\qe_1$-sound and $(c_0,\qe_2)$-complete.
\end{lemma}
\begin{corollary}
\label{corol:AnuL}
Let $\qa_2$ be parametrized as $\qa_2=\qo/c_0$, with $\qo<1$ a positive constant. Let the function $g_1(p)$ be such that it has the asymptotic behavior $g_1(p)\propto p^{-\qg}$ at $p=t$, with~$\qg< 1$. 
Then 
for large $c_0$, it is possible to achieve Soundness and Completeness with
$A=4\nu L^2[1+{\cal O}(1/c_0)]$.
\end{corollary}

{\it Proof of Lemma~\ref{lemma:AnuL}}:
The value of $c_0$ in (\ref{constructLc0}) is specifically set such that 
(\ref{ccondition1}) is satisfied, and hence 
we are allowed to use 
the inequality (\ref{conditionright}).
By combining the results (\ref{conditionright}) and (\ref{conditionleft}) 
we obtain a `window' for $A$ in which Soundness and Completeness are both satisfied,
\be
	A\in \left[LB+\frac{L}{\qa_2 c_0^2}\frac{\ln\qe_2}{\ln\qe_1},
	\quad \frac{B^2}{4\nu}\right].
\label{windowA}
\ee
We choose $B$ such that (i) this window exists, and (ii) the left boundary is as small as possible.
It turns out that the optimal choice for $B$ is the smallest value for which the window still exists. 
Setting the left and right boundary in (\ref{windowA}) equal to each other and solving for $B$ gives
\be
	B_{\rm opt}=4\nu L(1+\qj),
\label{BoptnuL}
\ee
with $\qj$ as defined in~(\ref{defqj}).
The corresponding value for $A$ is
$A_{\rm opt}=B_{\rm opt}^2/(4\nu)=4\nu L^2 (1+\qj)^2$.
\hfill$\square$

{\it Proof of Corollary~\ref{corol:AnuL}}:
First we show that the quantity $L>0$ is well defined 
in the limit of large~$c_0$.
To this end we inspect~(\ref{condition}), setting $c=c_0$.
We start with the term involving 
$\qD=e[\qo/1.7]^{1/(1-\qb)}<1$,
(with $\qo<1$ and $\frac{1}{1-\qb}>1$). 
The exponent $c_0-x_{\rm max}$, with $x_{\rm max}$ defined in (\ref{defxmax}), satisfies
\be
	c_0-x_{\rm max}\geq \qt
	+\frac{1.7 (1-t)c_0}{\qo g_1(\qt/c_0)}.
\ee
From the choice $g_1(t)\propto t^{-\qg}$, with $\qg< 1$, it follows that
$c_0-x_{\rm max}={\cal O}(c_0^{1-\qg})$, i.e. an increasing function of $c_0$. 
Hence the expression $\qD^{c_0-x_{\rm max}}/(1-\qD)$ decreases as 
$\exp[-\ql c_0^{1-\qg}]$ for some positive constant $\ql$. For large $c_0$, this is much faster than the term $\qa_2/L =\qo/c_0 L$.

Next we look at the term $\nu c_0\qa_2^2$. This can be written as $\qa_2 \nu\qo$. Hence, the factor multiplying $\qa_2$ is a small constant independent of~$c_0$. 

In the term $\qa_2 c_0\Ep[p^{c_0}g_1(p)]$, the factor multiplying
$\qa_2$ can be written in the form
\be
	c_0\Ep[p^{c_0}g_1(p)]=
	\left[ pf(p)g_1(p) p^{c_0}\right]_{p=\qt/c_0}^{1-\qt/c_0}
	-\int_{\qt/c_0}^{1-\qt/c_0}\!\! \rd p\; 
	p^{c_0} \frac{\rd}{\rd p}[pf(p)g_1(p)].
\ee
In this form, it is clear that the expression is finite for 
$c_0\naar\infty$, provided that the product $f(p)g_1(p)$ is non-pathological near~$p=1-t$. 

The remaining term in (\ref{condition}) is 
$\qa_2\cdot\sum_{x=1}^{c_0-1}{c_0\choose x}K_x^{\rm bound}$,
where $K_x^{\rm bound}$ is some nonnegative number that upper bounds the expression 
$K_{1,x}$ as defined in~(\ref{defK1x}).
From (\ref{defK1x}) it follows that
\be
	K_{1,x}\leq t f(t) g_1(t) t^x (1-t)^{c_0-x}
	+\int_J \!\!\rd p\; p^x (1-p)^{c_0-x}\frac{\rd}{\rd p}(pfg_1),
\label{ineqK1x}
\ee
where $J$ is defined as the interval (or set of intervals) on which
$\frac{\rd}{\rd p}(pfg_1)>0$. Using the expression (\ref{ineqK1x}) as our bound $K_x^{\rm bound}$ and computing the $x$-sum we obtain
\be
	\sum_{x=1}^{c_0-1}{c_0\choose x}K_x^{\rm bound}=
	tf(t)g_1(t)[1-(1-t)^{c_0}-t^{c_0}]
	+\int_J \!\!\rd p\; 
	[1-(1-p)^{c_0}-p^{c_0}]\frac{\rd}{\rd p}(pfg_1).
\label{sumKxbound}
\ee
Expressed in this form, it is clear that this contribution is of order
$1-(1-t)^{c_0}={\cal O}(\qt)$
for $c_0\gg 1$.
The smallness of all the expressions in (\ref{condition}) implies that
$L$ is well-defined.

The last step in the proof of Corollary~\ref{corol:AnuL} is the asymptotic behavior of the factor $(1+\qj)^2$ in (\ref{constructLAB}).
The fraction $1/(\qa_2 c_0^2)$ in the definition (\ref{defqj}) of $\qj$
is equal to $1/(\qo c_0)$, i.e. of order ${\cal O}(1/c_0)$.
Consequently, $(1+\qj)^2=1+{\cal O}(1/c_0)$.
\hfill $\square$

Note that the expression $4\nu L^2$ in Lemma~\ref{lemma:AnuL} originates from the specific proof technique.

The next step in the proof of Theorem~\ref{theorem:provablebound}
is to choose $f$ and $g_1$ such that the product $\nu L^2$ is minimized.
We refer to this choice as `optimal', but it should be clear that it represents optimality only with respect to the proof technique that we employ in Section~\ref{secreduceA}.


\subsection{Finding the optimal $g_1$ function}
\label{secgoptimal}
\begin{lemma}
\label{lemma:optimalg}
For all distributions $f$, 
Tardos' choice $g_1=g_1^{\rm T}$ is optimal in the sense that it minimizes 
the factor $\nu L^2$ appearing in the length parameter $A$ in Lemma~\ref{lemma:AnuL}.
This choice yields $\nu=1$.
\end{lemma}
{\it Proof}:
We have to solve an optimization problem and determine where the functional 
derivative of $4\nu L^2$ is zero.
This is easiest to accomplish by choosing as the independent degrees of freedom $f(p)$ and
$s(p):=pf(p)g_1(p)$ on the interval $p\in(t,1/2)$, instead of $f$ and~$g_1$.

From (\ref{condition}) it can be seen that $L$ depends on $f$ and $g_1$ only through~$s(p)$:
The expression $\Ep[p^c g_1]$ involves only the product $fg_1$; and
$K_x^{\rm bound}$ is a bound on (\ref{defK1x}), which also depends on $f$ and $g_1$ solely through the product $fg_1$.

The parameter $\nu$, on the other hand, depends on both the $s(p)$ and $f(p)$ degrees of freedom; Eq.(\ref{defnu})
can be rewritten as
\be
		 \nu[s,f]=2\int_t^{1/2}\!\!\!\!{\rm d}p\;
		 \frac{s^2(p)}{p(1-p)f(p)}.
\ee
The functional that we have to minimize is
\be
		 {\cal F}[s,f]=4\nu[s,f] L^2[s]
		 +\ql\left[ \int_t^{1/2}\!\!\! \rd p\; f(p) -1/2 \right],
\ee
where $\ql$ is a Lagrange multiplier for the normalization constraint on~$f$.
Setting the functional derivative with respect to $f(p)$ equal to zero gives
\bea
		 0=\frac{\qd{\cal F}}{\qd f(p)}=-8L^2\frac{s^2(p)}{p(1-p)f^2(p)}+\ql
		  & \hskip2mm\Longrightarrow\hskip2mm &
		 g_1(p)\propto \sqrt{\frac{1-p}{p}}.
\eea
Hence, since the normalization of $g_1$ is arbitrary\footnote{
As can be seen from the accusation rule (\ref{threshold}), rescaling $g_1$ and $Z$ by the same factor leaves the scheme invariant.}
, it turns out that Tardos' choice
$g_1(p)=g_1^{\rm T}(p)=\sqrt{(1-p)/p}$ is optimal.

From the optimal $g_1$ the value of $\nu$ follows directly, without dependence on~$f$: 
Substituting $g_1^{\rm T}$ into (\ref{defnu}) and using the fact that $f$ is a normalized probability distribution, we get $\nu=1$.\hfill$\square$


\subsection{The choice $f=f^{\rm T}$ seems to be optimal}
\label{secfoptimal}
Having found the optimal $g_1$, we can next search for the optimal distribution function~$f$.
The terms proportional to $\qa_2$ in the left hand side of (\ref{condition}) are the most important in determining the allowed values of $L$: we want to tune $f$ such that the number multiplying $\qa_2$ is as negative as possible.
We have looked at a class of smooth functions of the form
$f(p)=\half p^{a-1}(1-p)^{b-1}/[B_{1/2}(a,b)-B_t(a,b)]$ on the interval $p\in(t,1/2)$, where $B$ is the incomplete Beta function.
Numerical inspection shows that Tardos' choice $a=b=\half$ is the best choice for $a$, $b$.
Of course, as we have not investigated the full function space of $f$,
this does not prove that $f^{\rm T}$ is the best possible choice.

In the rest of this paper we will work with $f=f^{\rm T}$.


\subsection{Last step in the proof of Theorem~\ref{theorem:provablebound}}
\label{sec4pi2}

We are now finally in a position to prove Theorem~\ref{theorem:provablebound}.
We show that Theorem~\ref{theorem:provablebound} follows from Lemma~\ref{lemma:AnuL} in the special case $f=f^{\rm T}$ and 
$g_1=g_1^{\rm T}$. 
In this special case, we have $\nu=1$ as shown by Lemma~\ref{lemma:optimalg}. Hence the condition on $c_0$ (\ref{constructLc0}) reduces to~(\ref{mainresultc0}).
Furthermore, the choice $g_1=g_1^{\rm T}$ gives $\qb=\half$ 
(see App.~\ref{appTheorem2}), whereby $\qD$ reduces to $D$, as defined in (\ref{defD}).
Next we use the property 
$\frac{\rd}{\rd p}[pf^{\rm T}g_1^{\rm T}]=0$
to explicitly evaluate the $\Ep$ expectation and the
$K_x^{\rm bound}$ term in~(\ref{condition}).
We make use of expression (\ref{sumKxbound}) for $K_x^{\rm bound}$ and get
\bea
	c_0\Ep[p^{c_0}g_1(p)]=\frac{(1-t)^{c_0}-t^{c_0}}{\pi-4t'}
	& ; &
	\sum_{x=1}^{c_0-1}{c_0\choose x}K_x^{\rm bound} =
	\frac{1-(1-t)^{c_0}-t^{c_0}}{\pi-4t'}.
\eea
This allows us to rewrite (\ref{condition}) as
\be
	1-\qa_2\left\{ \frac{2(1-t)^{c_0}-1}{\pi-4t'}-\qo 
	-e^{1.7}\frac{c_0}{\qo(1-D)}
	D^{\qt+1.7\qo^{-1}\sqrt{\qt}\sqrt{c_0-\qt}}
	\right\}
	<1-\frac{\qa_2}{L}.
\label{condremake}
\ee
The quantity $\qd$ in Theorem~\ref{theorem:provablebound} is specifically chosen
such that $L=\pi/(1-\qd)$ satisfies~(\ref{condremake}).
The $\xi$ in Theorem~\ref{theorem:provablebound} is given by $\qj$ (\ref{defqj})
after the substitution $\nu=1$, $\qa_2 c_0=\qo$, $L=\pi/(1-\qd)$.
Thus, (\ref{constructLAB}) reduces to~(\ref{ABxiqd}).
\hfill $\square$


\section{Numerical evaluation}
\label{secnumerics}
In the preceding section, we proved a result for large coalition sizes.
In this section, we numerically investigate how quickly convergence to this behavior occurs.

\subsection{Method}

Given that $f^{\rm T}$ and $g_1^{\rm T}$ are the optimal functions, we determine the optimal values for the remaining parameters. For fixed $(c_0, \qe_1, \qe_2)$, our task is to find
$(t, \qa_1, \qa_2, L, B)$ such that we obtain the smallest possible~$A$.
We mean `smallest' in the sense that Soundness and Completeness can be proven using the technique in Section~\ref{secreduceA}, but without the assumption that $c_0$ is large.
The following constraints must be satisfied:
\begin{itemize}
\item
From (\ref{P1res}), (\ref{defmZ}) and the required property 
$\prob[{\rm false\;positive}]< \qe_1$
we get the constraint
\be
	-Ac_0^2\qa_1^2+Bc_0\qa_1 < 1.
\ee
\item
Similarly, from (\ref{P2res}), (\ref{defmZ}) and the required property
$\prob[{\rm false\;negative}]<\qe_2$ we get the constraint
\be
	(-B+A/L)c_0^2 \qa_2\ln\qe_1  < \ln\qe_2.
\ee
\item
In Section~\ref{secProp1} the parameter $\qa_1$ was introduced such that $\qa_1<1.7/g_1(t)$. This gives
\be
	\qa_1 < 1.7\sqrt{t/(1-t)}.
\ee
\item
In Appendix~\ref{appTheorem2}, it is assumed that $\qa_2$ is so small that the function 
$(1-p)^{c-x}\exp[\qa_2(cp-x)/\sqrt{p(1-p)}]$ is a decreasing function of $p$ in the vicinity of $p=1-t$ for all $x\in\{1,\ldots,c-1\}$.
Let's denote this function as $z(p)$.
Its derivative near $p=1-t$ is approximately given by 
$\rd z/\rd p\approx-z\cdot (c-x)(1-p)^{-1}[1-\frac{\qa_2}{2\sqrt{1-p}}]$.
Hence, in order to ensure a negative sign of the derivative, $\qa_2$ has to satisfy
\be
	\qa_2 \lessapprox 2\sqrt{t}.
\ee
\item
In condition (\ref{condition}) with $f=f^{\rm T}$, $g_1=g_1^{\rm T}$ 
the $K_x^{\rm bound}$ term is easily evaluated. The resulting condition is
\be
	1-\qa_2\frac{2(1-t)^{c_0}-1}{\pi-4t'}+c_0\qa_2^2+
	e^{1.7}\frac{D^{c_0-x_{\rm max}}}{1-D}< 1-\frac{\qa_2}{L}
\label{condTar}
\ee
with $D=e(c_0\qa_2/1.7)^2$ and $c_0-x_{\rm max}=\lceil
c_0 t+ 1.7\sqrt{t(1-t)}/\qa_2\rceil \geq 1$.
\end{itemize}
The complicated $\qa_2$-dependence of (\ref{condTar}), containing a term of the form $\qa_2^{1/\qa_2}$, prevents us from finding an optimum analytically.
Instead, we have searched for optimum parameter values numerically, using a randomized method following these steps:
\begin{enumerate}
\item
Choose a random $t$ uniformly from the interval $(0,\frac{1}{2c_0})$.
\item
Choose $\qa_1$ uniformly from
$(0, \min\left\{1.7\sqrt{t/(1-t)}, 
\frac{1}{c_0}\frac{2(1-t)^{c_0}-1}{\pi-4t'} \right\})$.
\item
Choose $L$ uniformly from 
$(\frac{\pi-4t'}{2(1-t)^{c_0}-1}, \frac{1}{c_0\qa_1})$.
\item
Find the largest possible value of $\qa_2$ in the interval
$\left(0,\min\left\{2\sqrt{t}, \frac{1}{c_0}\left[ \frac{2(1-t)^{c_0}-1}{\pi-4t'} -\frac{1}{L}
\right]\right\}\right)$
that satisfies condition (\ref{condTar}).
\item
Compute $A=\frac{L(c_0\qa_1)^{-1}}{(c_0\qa_1)^{-1}-L}\left[
(c_0\qa_1)^{-1}+\frac{1}{c_0^2\qa_2}\frac{\ln\qe_2}{\ln\qe_1}
\right]$.
\end{enumerate}
Note that the optimal choice of $B$ (in terms of achieving small $A$) follows from (\ref{windowA}). When the optimal value for $B$ is used, the interval (\ref{windowA}) consists of a single point seen in the last step.
We repeat steps 1--5 multiple times and select the set of parameter values that yield the lowest value of~$A$, i.e. the shortest code length. 

\subsection{Numerical results}
As the parameter $A$ depends on $\qe_1$ and $\qe_2$ only through the ratio of their logarithms, we define the 
parameter $R=\frac{\ln\qe_2}{\ln\qe_1}$ to display our results. This allows us to represent dependence on three 
parameters using only two variables, namely $A_{\rm best}=A_{\rm best}(c_0,R)$. As we are primarily concerned 
with content distribution applications, we have chosen $10^{-15}\leq \qe_1 \leq 10^{-9}$ 
and $0.1\leq\qe_2\leq 0.5$ as plausible values. This gives $0.02 \lessapprox R\lessapprox 0.1$. 
We further consider $10 \leq c_0 \leq 80$. 

We plot the best code length parameter $A$ as a function 
of $c_0$ (for constant $R$) and as a function of $R$ (for constant $c_0$) in Fig.~\ref{figc} and Fig.~\ref{figR}, respectively. Note that these figures are derived from the same dataset. In Appendix~\ref{apptable}, we give the corresponding values of the $t$ and $B$ parameters, which are necessary to implement the fingerprinting scheme.
The numerical results indicate that the result $A\approx4\pi^2\approx 39.5$
(see Corollary~\ref{corol:mainresult})
 seems to be approached for large $c_0$ and small $R$. Moreover, The $R$-dependence of $A$ is slightly sublinear. Note that there is not much variation in the value of $A$, only of the order of 15\%.

\begin{figure}
\parbox[t]{7cm}
  {
\scalebox{0.5}{\includegraphics[angle=0,clip=]{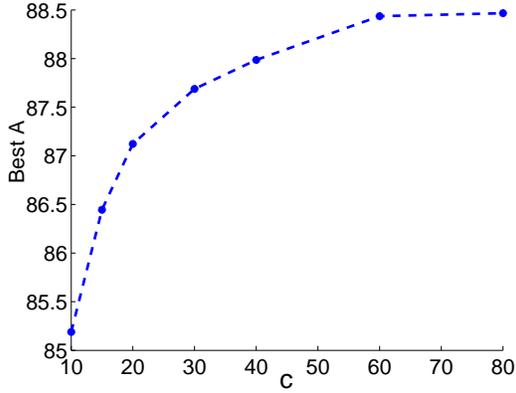}}
  \caption{\it Smallest obtained $A$ as a function of $c_0$ in the coupled case $R=c_0/4$.}
  \label{figcoupled}
  }
\end{figure}

In Fig.~\ref{figcoupled},
we also plot the results when the false positive and false negative rates are coupled as in~\cite{Tardos}, i.e. $\qe_2=\qe_1^{c_0/4}$ or $R=c_0/4$. In this case, it is possible to reduce $A$ to approximately 90, which is not much of an improvement with respect to Tardos' $A^{\rm T}=100$. This result further emphasizes 
the importance of decoupling FP and FN rates. When we allow for high FN rates $\qe_1 \ll \qe_2$, the code can be safely made more than a factor two shorter than suggested in Tardos' original construction.

Finally, the figures in this section, together with Appendix~\ref{apptable}, give a system parameter recipe
for content owners who wish to implement a provably $\qe_1$-sound and $(c_0,\qe_2)$-complete fingerprinting 
scheme with a code length $A(R,c_0)c_0^2 \ln\qe_1^{-1}$.
 
\begin{figure}[htb]
  \parbox[t]{7cm}
  {
  \scalebox{0.5}{\includegraphics[angle=0,clip=]{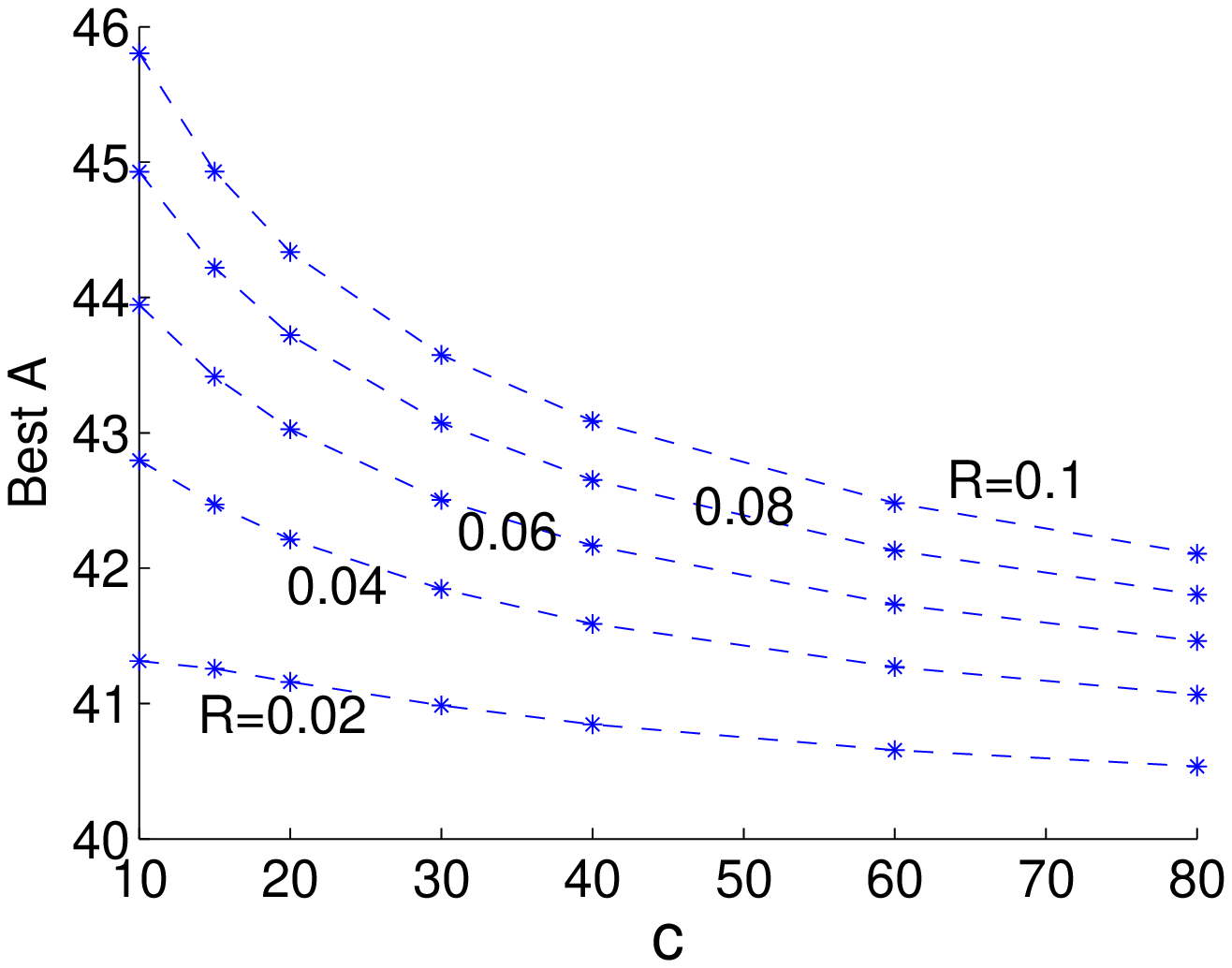}}
  \caption{\it Smallest obtained $A$ as a function of $c_0$ for various values of $R=\frac{\ln\qe_2}{\ln\qe_1}$.}
  \label{figc}
  }
  \hskip .5cm
  \parbox[t]{7cm}
  {
  \scalebox{0.5}{\includegraphics[angle=0,clip=]{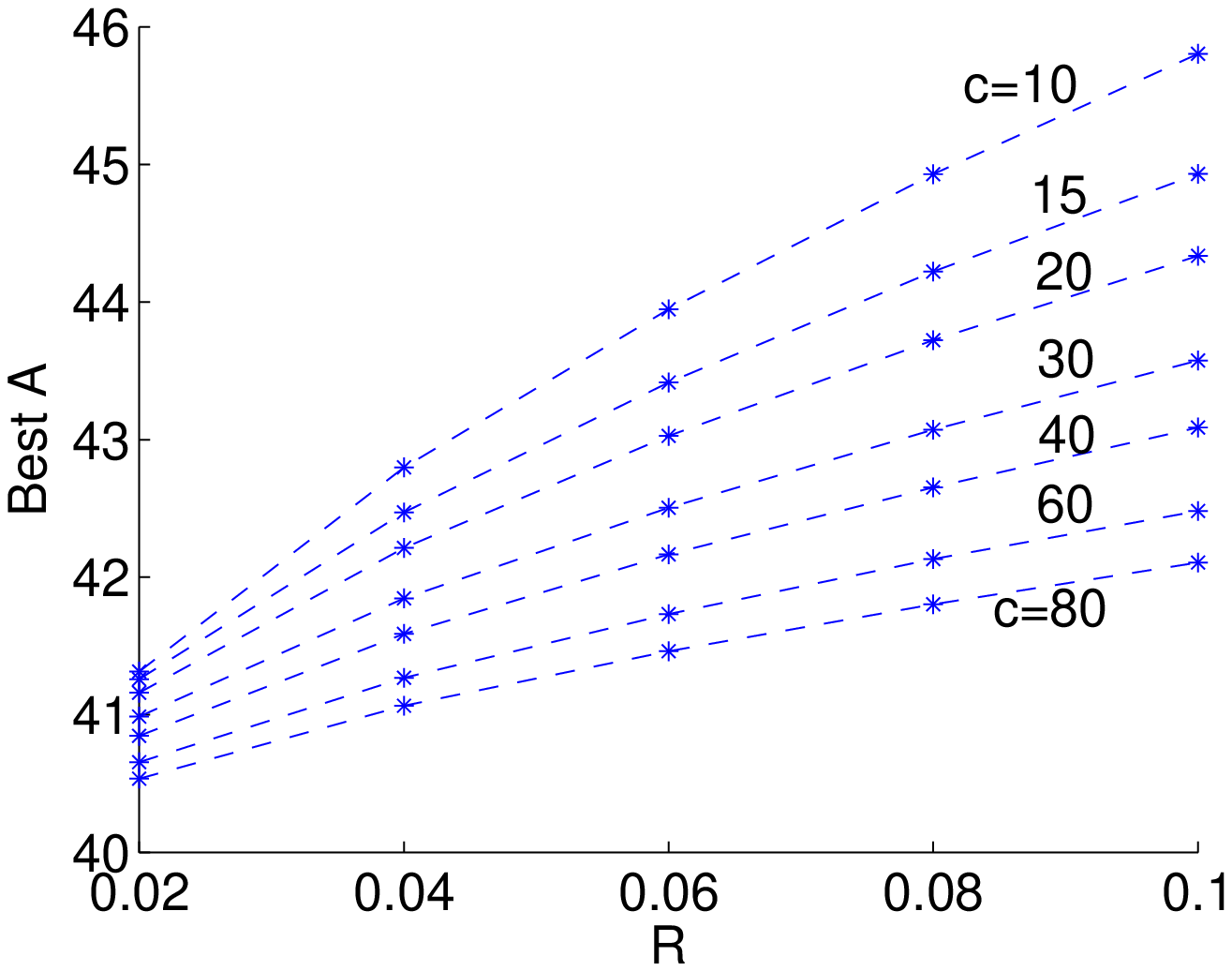}}
  \caption{\it Smallest obtained $A$ as a function of $R=\frac{\ln\qe_2}{\ln\qe_1}$ for various values of $c_0$.}
  \label{figR}
  }
\end{figure}

\section{Statistical approach}
\label{secstat}

\subsection{Motivation and main result of the statistical approach}
\label{statMainResult}
In this section, we put aside the provable properties of the Tardos fingerprinting scheme.
Instead we study the statistical behavior of the accusation sums $S_j$ (\ref{accuse}) for the innocent user 
and $S$ (\ref{defS}) for the coalition.
The advantage of the statistical approach is that we get more insight into the `true' 
behavior of the scheme (actual FP and FN probabilities as a function of $m$, $c_0$, $n$, $Z$) 
than provided by the provable result (Theorem~\ref{theorem:provablebound}) based on the Markov inequality. 

The accusation sums $S_j$ and $S$ are defined as the sum of a large number of stochastic variables. We expect
$S_j$ and $S$ to have an approximately Gaussian probability distribution.
(This is motivated in Appendix~\ref{appCLT}.)
The Central Limit Theorem states that a variable which is 
created by adding up many independent stochastic variables will have a 
Gaussian distribution in the vicinity of the distribution's peak.
The exact size of this `vicinity' depends on the distribution of the individual variables 
and on the total number of variables in the sum.
In Appendix~\ref{appCLT}, we argue that the Gaussian approximation 
is valid for some realistic values of $\qe_1$ and~$m$.

Our main result can be formulated as follows.
\begin{theorem}
\label{theorem:gaussian}
Let $\qe_1,\qe_2\in(0,1)$ be fixed parameters.
Let $c_0\in{\mathbb N}^+$ be a fixed parameter.
Let the functions $f(p)$ and $g_1(p)$ be given by $f=f^{\rm T}$, $g_1=g_1^{\rm T}$.
Let the cutoff parameter $t$ be parametrized as $t=\qt/c_0$.
Let the accusation sums $S_j$ (\ref{accuse}) and $S$ (\ref{defS}) obey Gaussian statistics. 
Then the fingerprinting scheme with code length $m$ and threshold $Z$ set according to
\bea
\label{mtheoremgauss}
	m &\geq& \frac{2\pi^2}{(1-2\qt)^2}\; c_0^2 \left[
	\Erfcinv(2\qe_1)+\frac{1}{\sqrt{c_0}}\Erfcinv(2\qe_2)
	\right]^2
	\\
	Z&\in&\left[\sqrt{2m}\;\Erfcinv(2\qe_1), \quad
	\frac{1-2\qt}{\pi c_0} m 
	- \frac{\sqrt{2m}}{\sqrt{c_0}} \Erfcinv(2\qe_2)
	\right]
\label{Ztheoremgauss}
\eea
is $\qe_1$-sound and $(c_0,\qe_2)$-complete.
\end{theorem}

Here Erfc stands for the complementary error function $1-\Erf$, with the definition 
$\Erf(x)=(2/\sqrt{\pi})\int_0^x \rd y\; e^{-y^2}$.
The superscript `inv' denotes the inverse function.

\begin{corollary}
\label{corol:gaussian}
Let $(\qe_1,\qe_2)\in(0,1)$ be independent fixed parameters.
Then for $c_0\gg 1$ the parameter choice 
\bea
	m=\frac{2\pi^2}{(1-2\qt)^2} c_0^2\ln\frac{1}{\qe_1\sqrt{2\pi}}
	&;&
	Z=\frac{2\pi}{1-2\qt} c_0 \ln\frac{1}{\qe_1\sqrt{2\pi}}
\label{mZcorol}
\eea
achieves $\qe_1$-soundness and $(c_0,\qe_2)$-completeness.
\end{corollary}

The proof of Theorem~\ref{theorem:gaussian}
is given in the coming sections and has the following outline.
First, in Sections \ref{secstatSj} and~\ref{secstatS}, we compute the lowest moments of the distributions of the accusation sums,
\bea
	\mu_j:=\EyXp[S_j] &; \hskip2mm& 
	\qs_j^2:=\EyXp[S_j^2]-\mu_j^2
	\hskip7mm {\rm where}\; j\;{\rm is\; not\; a\; colluder}
	\nn\\
	\mu:=\EyXp[S] &; \hskip2mm& \qs^2:=\EyXp[S^2]-\mu^2.
\label{stats}
\eea
Then, in Section~\ref{seccdep}, we compute the false positive and false negative error probabilities as a function of $m$, $Z$ and $c_0$.
We derive conditions on $m$ and $Z$ from the Soundness and Completeness requirements. 
In Section~\ref{secstrategy}, we identify an `extremal' strategy
which leads to a maximum value of $\qs_j$ and~$\qs$. 
In Section~\ref{secfinalgaussian}, we assume that the probability distributions are Gaussian
(a motivation for this step is given in Appendix~\ref{appCLT})
and we combine all the ingredients to complete
the final step in the proof of Theorem~\ref{theorem:gaussian}.
We prove Corollary~\ref{corol:gaussian} in Section~\ref{secproofcorol}.

\subsection{Statistics of an innocent user's accusation}
\label{secstatSj}
Even without knowing the colluders' strategy, we can derive 
a number of useful properties of
the expectation values listed in (\ref{stats}).
We start by looking at $S_j$, where user $j$ is not a colluder.
In Section~\ref{secnotation}, the functions $g_1, g_0$ were introduced such that $pg_1(p)+(1-p)g_0(p)=0$.
This immediately yields
\be
	\EE_{X_j}[S_j]=\sum_{i=1}^m y_i \;\EE_{X_{ji}}[U(X_{ji},p_i)]
	=0,
\label{expecU0}
\ee
where $y_i$ is shifted out of the expectation value because $j$ is not part of the coalition.
From (\ref{expecU0}) it immediately follows that $\mu_j=0$.

The standard deviation $\qs_j$ is computed as follows. Substitution of the definition 
(\ref{accuse}) into (\ref{stats}) gives
\be
	\qs_j^2 = \EyXp[S_j^2]=\sum_{i=1}^m \sum_{k=1}^m
	\EyXp[y_i y_k \; U(X_{ji},p_i)U(X_{jk},p_k)].
\ee
All terms with $i\neq k$ vanish, since then the expectation value factorizes into two 
parts that are both zero due to (\ref{expecU0}).
Hence we can write
\be
	\qs_j^2=\sum_{i=1}^m \EyXp[y_i\; U^2(X_{ji},p_i)]
	=\sum_{i=1}^m \EE_{yX_C p}\left[y_i \;
	\EE_{X_{ji}}[U^2(X_{ji},p_i)]\right].
\ee
(Here the expectation $\EE_{yX_C p}$ involves only those entries in $X$ that are visible to the colluders.)
Again we have used the fact that $y_i$ does not depend on $X_{ji}$ when user $j$ is innocent.
Next we make use of the property 
$\EE_{X_{ji}}[U^2(X_{ji},p_i)]=1$
which holds for $g_1=g_1^{\rm T}$.
This finally yields
\be
	\qs_j^2=\sum_{i=1}^m \EE_{yX_C p}[y_i]  <m.
\label{Sj2}
\ee

\subsection{Statistics of the coalition accusation $S$}
\label{secstatS}
Next we look at the collective accusation sum $S$ defined in (\ref{defS}). 
Now we have to keep in mind that $y_i$ depends on $X_{ji}$ when $j$ is a colluder.
Taking the expectation of (\ref{defS}) we get
\be
	\mu=\EyXp[S]=\sum_{i=1}^m \Ep\left[
	\EE_X\left[\EE_y[y_i]\{ x_i g_1(p_i)+[c-x_i]g_0(p_i) \}
	\right]\vphantom{M^M}\right].
\label{Sstatmu1}
\ee
The notation $\EE_y$ stands for the expectation value over the $y$ degrees of freedom for fixed $p$ and $X$.
In (\ref{Sstatmu1}) the expectation value over $X$, for fixed $p$, reduces 
to a binomial distribution on the integers~$x_i$ \cite{Tardos}:
\be
	\prob[\# \mbox{`1'} 
	{\rm \; entries\; in\; column}\; i\; {\rm is}\;x_i]
	={c\choose x_i}p_i^{x_i}(1-p_i)^{c-x_i}.
\label{defEXx}
\ee
We now evaluate (\ref{Sstatmu1}) as follows.
We express the $\EE_X$ expectation in the form (\ref{defEXx}).
We define a quantity $\qj_i(x_i):=\EE_{X_C p\backslash i}[\EE_y[y_i]]$.
Here the notation $\EE_{X_C p\backslash i}$ means the expectation value over all degrees of freedom
in $X_C$ and $p$ except column~$i$.
The quantity $\qj_i$ depends on $x_i$ and the colluder strategy (and possibly explicitly on $i$,
if the colluders choose to apply different strategies in different positions); 
it does not depend on $p_i$, as the colluders do not have access to $p_i$.
Finally we substitute Tardos' functions $g_1^{\rm T}$, $g_0^{\rm T}$ and $f^{\rm T}$. 
In this way we obtain, after some algebra,
\be
	\mu = \frac{1}{\pi-4t'}\sum_{i=1}^m
	\sum_{x_i=0}^c{c\choose x_i}\qj_i(x_i)
	\left\{ (1-t)^{x_i} t^{c-x_i}- t^{x_i}(1-t)^{c-x_i}
	\vphantom{\int}\right\}.
\label{mu1}
\ee
We now make use of the marking condition, giving
$\qj_i(0)=0$ and $\qj_i(c)=1$.
This allows us to rewrite (\ref{mu1}) as
\be
	\mu = \frac{1}{\pi-4t'}\sum_{i=1}^m\left[
	(1-t)^c-t^c 
	+\sum_{x_i=1}^{c-1}{c\choose x_i}\qj_i(x_i) \left\{\vphantom{\int}
	(1-t)^{x_i} t^{c-x_i}- t^{x_i}(1-t)^{c-x_i}
	\right\}\right].
\label{mu2}
\ee
Note that the coalition strategy has an almost negligible effect on~$\mu$.
The terms ${c\choose x}t^x(1-t)^{c-x}$ add up to 1 when the full sum is taken, but only the $x=0$ term is of order~1. All the other terms summed together are only of order $c_0 t=\qt\ll 1$. The same argument holds for the other summand, but there the $x=c$ term is dominant.

We use the same methods as above to evaluate $\qs$. Without showing the details of the computation, we give the result,
\be
	\qs^2=\frac{1}{\pi-4t'}\sum_{i=1}^m\sum_{x_i=1}^c 
	{c\choose x_i}\qj_i(x_i)
	\int_t^{1-t}\!\!\!\rd p\; (x_i-cp)^2 p^{x_i-3/2}(1-p)^{c-x_i-3/2}
	- \frac{\mu^2}{m}.
\label{S2}
\ee


\subsection{Relating $m$ and $Z$ to the error probabilities}
\label{seccdep}
In (\ref{Sj2},\ref{mu2},\ref{S2}) we see 
from the $i$-summations
that the $m$-dependence becomes very simple if the colluders apply the same
strategy in each column;
namely, the quantities $\qs_j^2$, $\mu$ and $\qs^2$ then all become proportional
to~$m$. This motivates us to define `scaled' quantities as follows,
\bea
	\qs_j^2=m\tilde\qs_j^2
	& \hskip3mm
	\mu = m\tilde \mu
	& \hskip3mm
	\qs^2 = m \tilde\qs^2.
\label{mscale}
\eea
Let us introduce the notation 
$\qr_1$ and $\qr_2$
for the probability distribution functions of $S_j$ and $S$, respectively. 
These functions are unknown to us, but we normalize them so that they have zero mean and unit variance, 
\bea
	\prob[S_j\in[s,s+\tri s] ]= 
	\qr_1(\frac{s}{\qs_j})\frac{\tri s}{\qs_j}
	&; \hskip2mm&
	\prob[S\in[s,s+\tri s] ]=
	\qr_2(\frac{s-\mu}{\qs})\frac{\tri s}{\qs},
\label{defrho}
\eea
with $\int_{-\infty}^\infty \rd x\; \qr_1(x)=1$
and $\int_{-\infty}^\infty \rd x\; \qr_2(x)=1$.
We introduce the cumulative `tail' functions as
\bea
	G_1(x)=\int_x^\infty \!\! \rd x' \qr_1(x')
	& ; \hskip 3mm&
	G_2(x)=\int_{-\infty}^x \!\! \rd x' \qr_2(x').
\label{defG1G2}
\eea
With this notation, the error probabilities are expressed as
\bea
	\mbox{FP error prob.}=G_1(\frac{Z}{\qs_j})
	&  \hskip2mm; \hskip2mm &
	\mbox{FN error prob.}= G_2(\frac{c_0 Z-\mu}{\qs}).
\eea
\begin{figure}
\begin{center}
\includegraphics[width=10cm]{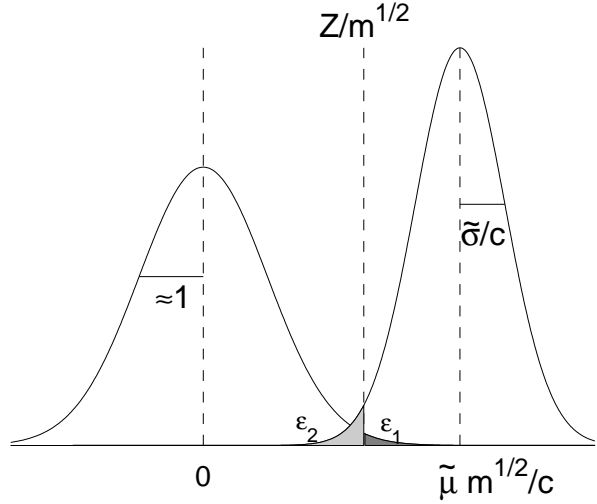}
\caption{Sketch of the probability density of $S_j/\sqrt{m}$ (left)
and $\frac{1}{c}S/\sqrt{m}$ (right). 
The accusation threshold $Z$ and the desired error rates $\qe_1$ and $\qe_2$ are also shown.}
\label{fig:gaussians}
\end{center}
\end{figure}
This is sketched in Fig.~\ref{fig:gaussians}.
The left curve is the probability density of the quantity $S_j/\sqrt{m}$.
It has mean $\tilde\mu_j=0$ and variance $\tilde\qs_j\approx 1$ (this will be shown in Section~\ref{secstrategy}).
The FP error rate (which should be less than $\qe_1$) is given by the area to the right of the (rescaled) threshold
$Z/\sqrt{m}$.
The right curve is the probability density of the quantity
$\frac{1}{c}S/\sqrt{m}$.
It has average $\frac{1}{c}\tilde\mu\sqrt{m}$ and variance $\tilde\qs/c$.
The FN error rate is given by the area to the left of
$Z/\sqrt{m}$.
The horizontal axis is scaled such that the $S_j$-curve does not depend on $c$ and~$m$. Several important properties follow from this picture:
\begin{itemize}
\item
For fixed $m$ and $Z$, increasing $c$ beyond $c_0$ affects only the 
distribution of~$S$. The left curve remains unchanged and hence the FP rate is independent of the coalition size. This is compatible with 
the definition of $\qe_1$-soundness (Definition~\ref{def:soundness}).
\item
The FP error rate is determined by one parameter: $Z/\sqrt{m}$.
Hence $Z$ must be chosen as $Z\propto\sqrt{m}$ as far as the dependence on $c_0$ is concerned.
\item
When $c$ increases, the rightmost curve becomes narrower and shifts to the left. 
In order to prevent the center of this curve ($\propto \sqrt{m}/c$) from crossing the threshold line, we need $m=\qO(c^2)$.
Together with the previous point,
this illustrates the need for the proportionalities $m\propto c_0^2$,
$Z\propto c_0$ in the Tardos scheme.
\end{itemize}
More precise results are derived next.
\begin{lemma}
\label{lemma:Zinterval}
A sufficient condition for $\qe_1$-soundness and $(c_0,\qe_2)$-completeness is given by
\be
	Z\in\left[\tilde\qs_j \sqrt{m}\; G_1^{\rm inv}(\qe_1),\quad
	\frac{\tilde\mu}{c_0}m+\frac{\tilde\qs}{c_0}\sqrt{m}\;
	G_2^{\rm inv}(\qe_2)  \right].
\label{ZregionG}
\ee
\end{lemma}
{\it Proof of Lemma~\ref{lemma:Zinterval}}:
The left boundary directly follows from the requirement 
$G_1(Z/\qs_j)\leq\qe_1$, using the notation~(\ref{mscale}). 
The right boundary follows from the requirement
$G_2([c_0 Z-\mu]/\qs)\leq\qe_2$.\hfill $\square$

Note that $G_2^{\rm inv}(\qe_2)$ is negative on the interval 
$\qe_2< 1/2$ for symmetric $G_2$, 
and that it monotonically increases as a function of $\qe_2$.

The $Z$-interval (\ref{ZregionG}) exists only for sufficiently large~$m$. 
One can think of a region in the $(u,v)$-plane, with $u=\sqrt{m}$ and $v=Z$, bounded on the lower side by a line $v\propto u$ and on the upper side by a quadratic function of~$u$.
The linear and quadratic curve meet each other at $m=m_{\rm min}$,
\be
	m_{\rm min}=\tilde\mu^{-2} c_0^2 \left[
	\tilde\qs_j\; G_1^{\rm inv}(\qe_1)-
	\frac{\tilde\qs}{c_0}\; G_2^{\rm inv}(\qe_2)
	\right]^2.
\label{mmin}
\ee
This represents the smallest possible code length for which the interval
(\ref{ZregionG}) exists, and hence, by Lemma~\ref{lemma:Zinterval}, 
the smallest possible code length for which the code is properly collusion resistant.

\subsection{`Extremal' colluder strategy}
\label{secstrategy}

Eq.~(\ref{mmin}) allows us to find the `worst case' or `extremal' colluder strategy. 
We define this as the strategy that causes the highest possible value of $m_{\rm min}$. 
Even though the colluders do not necessarily use this strategy, the content owner has 
to take into account that they {\em might} and has to adjust $m$ accordingly.
We make the following observations:
\begin{itemize}
\item
One way of increasing $m_{\rm min}$ would be to make $\tilde\mu$ as small as possible. 
However, in Section~\ref{secstatS} it was shown that the choice of strategy has negligible effect on~$\tilde\mu$.
\item
Another way of increasing $m_{\rm min}$ is to make $\tilde\qs_j$ and $\tilde\qs$ as large as possible. 
Here the choice of strategy has a big impact.
Both $\tilde\qs_j$ and $\tilde\qs$ are maximally large if the coalition outputs a `1' whenever possible.
\end{itemize}
We see that the `extremal' strategy is to set $\qj_i(x_i)=1$ for $x_i\neq 0$.
(The marking condition enforces $\qj_i(0)=0$.)
It looks as if the colluders are incriminating themselves in those columns where $p_i<1/2$.
However, for each symbol they equally incriminate a fraction $p_i$ of all the other users. 
The strategy derives its effectiveness from the large number of users who get accused along with the colluders.

Substitution of the extremal strategy into (\ref{Sj2}) and (\ref{S2}) with $c=c_0$ gives
\bea
	\tilde\qs_j^2 &=& 
	1-\frac{1}{\pi-4t'}\int_t^{1-t}\!\!\!\rd p\; 
	p^{-1/2}(1-p)^{c_0-1/2}
	= 1-\frac{1}{\sqrt{\pi c_0}}+{\cal O}(c_0^{-1})
	\nn\\
	\tilde\qs^2 +\tilde\mu^2 &=& 
	c_0-\frac{c_0^2}{\pi-4t'}
	\int_t^{1-t}\!\!\!\rd p\; p^{1/2}(1-p)^{c_0-3/2}
	= c_0[1-\frac{1}{2\sqrt{\pi c_0}}
	+{\cal O}(\frac{1}{c_0})]
\label{yconseq}
\eea

\subsection{Final step in the proof of Theorem~\ref{theorem:gaussian}}
\label{secfinalgaussian}

If $S_j$ and $S$ have a Gaussian distribution, then the functions $G_1$, $G_2$ become error functions, and we have
\bea
	G_1^{\rm inv}(\qe_1)=\sqrt{2}\;\Erfcinv(2\qe_1)
	&\quad;\quad&
	G_2^{\rm inv}(\qe_2)=-\sqrt{2}\;\Erfcinv(2\qe_2).
\label{GErfcinv}
\eea
We obtain the following inequalities from (\ref{yconseq}),
\bea
	\tilde\qs_j <1,
	\quad &;& \quad
	\tilde\qs < \sqrt{c_0}.
\label{ineq2}
\eea
These are independent of the choice of strategy function $\qj_i$.
Likewise, from (\ref{mu2}) we also obtain an inequality that is independent of
$\qj_i$.
This is done by taking only the negative part of the summand in (\ref{mu2}) and
setting $\qj_i=1$. The result is
\be
	\tilde\mu > \frac{1-2\qt}{\pi}.
\label{ineqmu}
\ee
We substitute (\ref{GErfcinv}) into (\ref{ZregionG}) and (\ref{mmin}),
and then use the inequalities (\ref{ineq2},\ref{ineqmu}).
This exercise shows that the choice (\ref{mtheoremgauss}) for $m$
is larger than $m_{\rm min}$, as it should indeed be, 
and that the $Z$-interval (\ref{Ztheoremgauss}) lies within the interval~(\ref{ZregionG}).
This completes the proof of Theorem~\ref{theorem:gaussian}.
\hfill$\square$


\subsection{Proof of Corollary \ref{corol:gaussian}}
\label{secproofcorol}
We use the inequality \cite{Wolfram}
\be
	\ln\frac{1}{x}\sqrt{\frac{2}{\pi}} > 
	\left[ \Erfcinv x \right]^2
\ee
to prove that the expression $\ln(1/\qe_1\sqrt{2\pi})$
in (\ref{mZcorol})
is larger than $[\Erfcinv(2\qe_1)]^2$.
Thus the following code length achieves Soundness and Completeness,
\be
	m= \frac{2\pi^2}{(1-2\qt)^2}c_0^2 \ln\frac{1}{\qe_1\sqrt{2\pi}}
	\left[
	1+\frac{1}{\sqrt{c_0}}\sqrt{\frac{\ln\qe_2\sqrt{2\pi}}
	{\ln\qe_1\sqrt{2\pi}}}
	\right]^2.
\label{lnlnfrac}
\ee
We then neglect the term containing $\qe_2$ with respect to 1, since it is of order ${\cal O}(1/\sqrt{c_0})$.
This yields the value of $m$ in Corollary \ref{corol:gaussian}.
Finally,
the value of $Z$ in Corollary \ref{corol:gaussian} follows by substituting this $m$ into the left boundary in~(\ref{Ztheoremgauss}).
\hfill $\square$

{\it Remarks}:
In the regime $\qe_1\ll \qe_2$, the fraction of logarithms in (\ref{lnlnfrac}) is typically smaller than~0.1 (see Section~\ref{secnumerics}).
Hence, the asymptotic result 
is already approached for relatively small values of~$c_0$.

In the case where $\qe_2$ and $\qe_1$ are coupled according to $\qe_2=\qe_1^{c_0/4}$, as was done in Tardos' original construction, Corollary~\ref{corol:gaussian} does not hold, as
we get $\ln \qe_2=(c_0/4)\ln\qe_1$. Here the fraction of logarithms is not negligible and leads to a factor $(9/2)\pi^2$ in $m$ instead of $2\pi^2$.


\section{Summary}

We have reevaluated the performance of the Tardos fingerprinting scheme by parameterizing its numerical 
constants and fixed functions. We have further modified the scheme by decoupling the desired false negative and 
false positive error probabilities. Using a proof technique similar to the one in~\cite{Tardos}, we 
have shown how short the code length can be with provable $\qe_1$-soundness and $(c_0,\qe_2)$-completeness. 
The main results of our study can be summarized as follows:
\begin{itemize}
\item
Tardos' accusation function $g_1^{\rm T}$ is `optimal' in the sense that it minimizes the provably sufficient
code length for our particular choice of proof method.
\item
Tardos' probability distribution function $f(p)$ is `optimal' in the same sense
within a limited class of functions which has the form $p^{a-1}(1-p)^{b-1}$.
\item
For sufficiently large $c_0$ values, and $\qe_2$ independent of $\qe_1$, the code length can be 
reduced from Tardos' $100 c_0^2 \ln\qe_1^{-1}$ to 
approximately $4\pi^2 c_0^2 \ln\qe_1^{-1}$. 
\item
When $\qe_2\gg \qe_1$, for instance for content distribution applications, our numerical results show that a code length
$m < 46  c_0^2 \ln\qe_1^{-1}$ is achievable already for $c_0 > 9$.
\item 
For sufficiently large $c_0\gg 1$, the accusation sums $S_j$ for the innocent user and $S$ for the coalition 
have probability distributions which are very close to Gaussian---due to the Central Limit Theorem.
If these distributions are perfectly Gaussian, then,
in the case of independent $\qe_1$, $\qe_2$,
a code length of $m\approx 2\pi^2 c_0^2\ln\qe_1^{-1}$ is sufficient for achieving $\qe_1$-soundness and $(c_0,\qe_2)$-completeness.
\end{itemize}

\appendix

\section{Condition for Completeness}
\label{appTheorem2}

In this appendix we derive an upper bound on the expression
$\EyXp[e^{-\qa_2 S}]$.
The first part of the derivation is directly copied from~\cite{Tardos}, so we will not repeat it here. 
We start our analysis at the earliest point where the approach with general $A$, $B$, $f$ and $g$ deviates from~\cite{Tardos}.

From partial evaluation of the $X$-average (which involves the binomial distribution for each column of $X$ separately) and from $|i: y_i=1| \leq m$, it can be shown that
\bea
\label{summaxE}
	\EyXp[e^{-\qa_2 S}] & \leq &
	\left[
	E_{0,0}+E_{1,c}+\sum_{x=1}^{c-1}{c \choose x}\max(E_{0,x},E_{1,x})
	\right]^m
	\\
	E_{0,x} &:=& \Ep[p^x(1-p)^{c-x}] \nn\\
	E_{1,x} &:=& \Ep\left[p^x(1-p)^{c-x}\exp\left(-\qa_2\{
	xg_1(p)+[c-x]g_0(p)\}\vphantom{M^{T^2}}\right)\right]
	\nn\\ &=&
	\Ep\left[p^x(1-p)^{c-x}\exp \left(\qa_2 g_1(p)\frac{cp-x}{1-p}
	\right)\right].\nn
\eea
The term $E_{1,c}$ is easily bounded,
\be
	E_{1,c}=\Ep[p^c e^{-c\qa_2 g_1}]\leq E_{0,c}-c\qa_2
	\Ep[p^cg_1]+c^2\qa_2^2 \Ep[p^c g_1^2].
\label{E1cbound}
\ee
Next it is proven that (for $x=1\ldots c-1$) $\max(E_{1,x},E_{0,x})\leq E_{0,x}+$ some positive expression.
To this end the inequality $e^u\leq 1+u+u^2$ is again used, which holds for $u<1.7$.
\bea
	&& p^x(1-p)^{c-x}\exp \left(\qa_2 g_1(p)\frac{cp-x}{1-p}\right)
	 \leq 
	p^x(1-p)^{c-x}\left\{
	1+\qa_2 g_1(p)\frac{cp-x}{1-p}
	+\qa_2^2\left[g_1(p)\frac{cp-x}{1-p}\right]^2
	\right\} \nn\\
	& & + \qY\left(\qa_2 g_1(p)\frac{cp-x}{1-p}-1.7\right)
	(1-p)^{c-x}\exp \left[\qa_2 g_1(p)\frac{cp-x}{1-p}\right].
\label{boundexp}
\eea
Here the term with the step function $\qY$ ensures that the right-hand side is always larger than the left-hand side, even if the expression in the exponent exceeds~1.7, which may happen for $p\in(x/c,1-t)$ if $\qa_2$ is not very small.
Taking the expectation value of (\ref{boundexp}), we get
\be
	E_{1,x} \leq  E_{0,x}+\qa_2 K_{1,x}+\qa_2^2 K_{2,x}+R_x
\ee
with
\bea
	K_{1,x}& := & \Ep\left[
	p^x(1-p)^{c-x}g_1\frac{cp-x}{1-p}\right] \nn\\ &=&
	\left. \vphantom{\int}-pfg_1\cdot p^x(1-p)^{c-x} \right|_{p=t}^{1-t}
	+ \int_t^{1-t}\!\!\! \rd p\; p^x(1-p)^{c-x}\frac{\rd}{\rd p}(pfg_1)
\label{defK1x}
	\\
	K_{2,x} &:= &\Ep\left[
	p^x(1-p)^{c-x}\left\{g_1(p)\frac{cp-x}{1-p}\right\}^2\right]\geq 0
	\\
	R_x &:=&\Ep\left[
	\qY\left(\qa_2 g_1(p)\frac{cp-x}{1-p}-1.7\right)
	(1-p)^{c-x}\exp \left[\qa_2 g_1(p)\frac{cp-x}{1-p}\right]
	\right]\geq 0.
\label{defRx}
\eea
Now we have to upper bound $K_{1,x}$ by a nonnegative expression.
Here we depart from \cite{Tardos}.
Tardos makes a very specific choice for the $f$ and $g_1$ function, namely $p f(p) g_1(p)=$constant.
We keep the derivation as general as we can.
For the moment we simply assume that we can find tight bounds $K_x^{\rm bound}\geq 0$
such that
\be
		 K_{1,x}\leq K_x^{\rm bound}.
\ee
Then we have
\be
		 \max(E_{0,x},E_{1,x})\leq E_{0,x}+\qa_2 K_x^{\rm bound}
		 +\qa_2^2 K_{2,x}+R_x.
\label{maxbound}
\ee
Substituting (\ref{maxbound}) and (\ref{E1cbound}) into (\ref{summaxE}) we get
\bea
	\EyXp[e^{-\qa_2 S}] &\leq& \left[
	\sum_{x=0}^c{c\choose x}E_{0,x}
	-c\qa_2 \Ep[p^c g_1]
	+ \qa_2\sum_{x=1}^{c-1}{c\choose x}K_x^{\rm bound}
	+\qa_2^2\sum_{x=1}^c {c\choose x}K_{2,x}
	\right.\nn\\ && \left.
	+\sum_{x=1}^{c-1} {c\choose x}R_x
	\right]^m.
\label{overall1}
\eea
The $E_{0,x}$ term contains a sum over the binomial distribution and simply yields $\Ep[1]=1$.
The $K_{2,x}$ term is bounded as follows,
\be
	\sum_{x=1}^c {c\choose x}K_{2,x}\leq  
	\Ep\left[\frac{g_1^2}{(1-p)^2}
	\sum_{x=0}^c{c\choose x}p^x(1-p)^{c-x}(x-cp)^2\right]
	=c\Ep[g_1^2\frac{p}{1-p}]=c\nu.
\ee
Next we bound the $R_x$ term.
Note that the step function in (\ref{defRx}) for fixed $p$ is nonzero only if $x\leq x_{\rm max}$, where
\be
	x_{\rm max}:=\left\lfloor c(1-t)-1.7
	\frac{1-t}{\qa_2 g_1(t)} \right\rfloor
	<c.
\label{defxmax}
\ee
Furthermore, we note that the function multiplying the step function in (\ref{defRx}) is monotonically 
decreasing as a function of $p$, provided that $\qa_2$ is `small enough'.
(This statement is made more accurate in Section~\ref{secnumerics}).
This means that the expectation value $\Ep[\cdot]$ can be upper bounded by evaluating the integrand at 
the point $p=p^*_x$, the smallest value of $p$ where the step function is nonzero,
\bea
		 R_x \leq (1-p^*_x)^{c-x}
		 \exp\left[\qa_2 (cp^*_x-x)\frac{g_1(p^*_x)}{1-p^*_x}\right]
		 & \hskip5mm {\rm with} \hskip5mm&
		 \qa_2 (cp^*_x-x)\frac{g_1(p^*_x)}{1-p^*_x}=1.7.
\label{boundRx1}
\eea
We introduce a numerical constant $\qb$ such that
$g_1(p^*_x)\geq (1-p^*_x)^\qb$. (In Tardos' case $\qb=1/2$).
From the definition of $p^*_x$ (\ref{boundRx1}) it then follows that
\be
		 (1-p^*_x)\leq \left[\frac{\qa_2}{1.7}(cp^*_x-x)
		 \right]^{1/(1-\qb)}
		 < \left[\frac{\qa_2}{1.7}(c-x)
		 \right]^{1/(1-\qb)}.
\label{Rxbound2}
\ee
This gives us the following bound on $R_x$ for $x<c$:
\be
		 R_x < e^{1.7}\left[\frac{\qa_2}{1.7}(c-x)
		 \right]^{\frac{c-x}{1-\qb}}.
\label{Rxbound3}
\ee
The $R_x$-sum in (\ref{overall1}) can then be bounded as
\bea
	\sum_{x=1}^{c-1}{c\choose x}R_x &\leq&
	\sum_{x=0}^{c-1} {c\choose x}R_x  <  e^{1.7}
	\sum_{x=0}^{x_{\rm max}} {c\choose x}\left[\frac{\qa_2}{1.7}(c-x)
	\right]^{\frac{c-x}{1-\qb}}
	\leq e^{1.7}\sum_{x=0}^{x_{\rm max}} (\frac{ce}{c-x})^{c-x}
	\left[\frac{\qa_2}{1.7}(c-x)
	\right]^{\frac{c-x}{1-\qb}}
	\nn\\ &=&
	e^{1.7}\sum_{x=c-x_{\rm max}}^c (\frac{ce}{x})^{x}
	\left[\frac{\qa_2}{1.7}x\right]^{\frac{x}{1-\qb}}
	< e^{1.7}\sum_{x=c-x_{\rm max}}^c
	\left[e(\qa_2 c/1.7)^{\frac{1}{1-\qb}}\right]^x
	\nn\\ &=&
	e^{1.7} \left[e(\qa_2 c/1.7)^{\frac{1}{1-\qb}}
	\right]^{c-x_{\rm max}}\frac{1-\left[e(\qa_2 c/1.7)^{\frac{1}{1-\qb}}\right]^{x_{\rm max}+1}}{1-\left[e(\qa_2 c/1.7)^{\frac{1}{1-\qb}}\right]}
	\nn\\ & =: &
	e^{1.7}\qD^{c-x_{\rm max}}\frac{1-\qD^{x_{\rm max}+1}}{1-\qD}
	< e^{1.7}\qD^{c-x_{\rm max}}\frac{1}{1-\qD}
\label{Rxbound4}
\eea
where we have introduced the abbreviation $\qD$ for the small\footnote{
As long as $\qb$ does not deviate too much from the Tardos case, $\qD$
is of order $>1$ in the small parameter~$\qa_2$.}
value
$e(\qa_2 c/1.7)^{1/(1-\qb)}$.
Summarizing, from (\ref{overall1}) we obtain
\be
	\EyXp[e^{-\qa_2 S}]<\left[1-c\qa_2 \Ep[p^c g_1]
	 + \qa_2\sum_{x=1}^{c-1}{c\choose x}K_x^{\rm bound}+\nu c\qa_2^2
	+e^{1.7}\qD^{c-x_{\rm max}}\frac{1}{1-\qd}\right]^m.
\ee
Finally we impose the following condition on the parameters $t$, $\qa_2$:
\be
	1-c\qa_2 \Ep[p^c g_1]
	 + \qa_2\sum_{x=1}^{c-1}{c\choose x}K_x^{\rm bound}+\nu c\qa_2^2
	+e^{1.7}\qD^{c-x_{\rm max}}\frac{1}{1-\qD} < 1-\frac{\qa_2}{L},
\label{condition}
\ee
where $L>0$ is a numerical constant. 
The satisfiability of this condition depends on
the choice of $f$, $g_1$ and~$L$.
Given that the condition is satisfied, we have the upper bound
\be
		 \EyXp[e^{-\qa_2 S}]<[1-\qa_2/L]^m \leq \exp (-\qa_2\frac{m}{L}).
\ee
For given $f$ and $g_1$, we will be interested in the smallest value of
$L$ that can be achieved.
Tardos chose $t$ and $\qa_2$ such that $L=4$.


\section{Numerical results}
\label{apptable}

Table~1 shows the results of the numerical experiments described in Section~\ref{secnumerics}.
The values $A$, $B$ and $t$ parametrize the code length, accusation threshold and $p$-axis cutoff, respectively. We also note, for the sake of completeness, that the auxiliary variables take the following values 
for the parameter sets indicated in the table: 
$1.001 \leq L/\pi \leq 1.026$,
$1.48 \leq \qa_1/\qa_1^{\rm T}\leq 1.58$ and
$0.05 \leq\qa_2/\qa_2^{\rm T}\leq 0.30$. 

\vskip4mm

\begin{center}
\begin{tabular}{|l||l|r|r|r|r|r|r|r|}\hline
\multicolumn{9}{|c|}{{\bf Table~1:} 
Numerical results
} \\ \hline
$R\downarrow$  & $\rightarrow c_0$& 10 & 15 & 20 & 30 & 40 & 60 & 80
\\ \hline
0.02 & $A$ & 41.31 & 41.26 & 41.16 & 40.99 & 40.85 & 40.66 & 40.54
    \\ \hline
    & $B$ & 12.86 & 12.85 & 12.83 & 12.80 & 12.78 & 12.75 & 12.73
    \\ \hline
    & $t/t^{\rm T}$ & 3.26 & 2.25 & 1.72 & 1.00 & 0.88 & 0.50 & 0.27
    \\ \hline
0.04 & $A$ & 42.80 & 42.47 & 42.21 & 41.85 & 41.59 & 41.27 & 41.06
    \\ \hline
    & $B$ & 13.08 & 13.03 & 13.00  & 12.94 & 12.90 & 12.85 & 12.82
    \\ \hline
    & $t/t^{\rm T}$ & 2.96 & 2.17 & 1.44 & 1.14 & 0.77 & 0.52 & 0.43
    \\ \hline
0.06 & $A$ & 43.95 & 43.41 & 43.03 & 42.50 & 42.17 & 41.73 & 41.46
    \\ \hline
    & $B$ & 13.26 & 13.18 & 13.12 & 13.04 & 12.99 & 12.92 & 12.88
    \\ \hline
    & $t/t^{\rm T}$ & 3.41 & 2.28 & 1.53 & 1.04 & 0.65 & 0.49 & 0.37
    \\ \hline
0.08 & $A$ & 44.93 & 44.22 & 43.72 & 43.07 & 42.65 & 42.13 & 41.80
    \\ \hline
    & $B$ & 13.41 & 13.30 & 13.22 & 13.13 & 13.06 & 12.98 & 12.93
    \\ \hline
    & $t/t^{\rm T}$ & 3.16 & 2.27 & 1.71 & 1.04 & 0.77 & 0.62 & 0.33
    \\ \hline
0.10 & $A$ & 45.80 & 44.93 & 44.34 & 43.58 & 43.09 & 42.48 & 42.11
    \\ \hline
    & $B$ & 13.54 & 13.41 & 13.32 & 13.20 & 13.13 & 13.04 & 12.98 
    \\ \hline
    & $t/t^{\rm T}$ & 3.31 & 1.94 & 1.60 & 1.23 & 0.77 & 0.46 & 0.39
    \\ \hline
\end{tabular}
\end{center}


\section{The Gaussian approximation}
\label{appCLT}

Under some reasonable assumptions, we can regard the accusation sums as Gaussian-distributed stochastic variables. Here, we outline our assumptions and show that the Central Limit Theorem (CLT) is applicable under these conditions. We first note the complete column symmetry and column independence of both the code generation process and the accusation method. Given this symmetry, we argue (without providing a proof) that the best colluder strategy for generating the colluded copy is also symmetric, i.e.  their output $y_i$ is independent of the
column index $i$ and independent of the $X_C$ entries in the other columns ($\neq i$). Note that the `extremal' colluder strategy of Section~\ref{secstrategy} also complies with this assumption.
Given column symmetry and mutual independence of the accusation values, under the assumption of a symmetric colluder strategy, the accusation sums $S$,  $S_j$ are sums of i.i.d. variables, and the Central
Limit Theorem (CLT) is applicable.

We show that the domain of applicability of the CLT is large enough to encompass a sufficient part of the tail of the $S_j$ and $S$ distributions, so that the approximations made in Section~\ref{secfinalgaussian} are justified.
The error probability $\qe_1\geq 10^{-15}$ represents at most an `8-sigma' event, i.e. we are interested in the region of 8 standard deviations $\qs_j$ around the average of~$S_j$.

First we determine the probability distribution of each separate accusation $U(X_{ji},p_i)$, for innocent $j$, given that $y_i=1$.
We define, for infinitesimal $\tri u$,
\be
	\prob[u\leq U \leq u+\tri u] = \qf(u)\tri u.
\ee
We compute the conditional probability that $U=u$ given $X_{ji}=1$.
We write
$\qf(u|X=1)\rd u = f(p)\rd p$, from which it follows that
$\qf(u|X=1)=f(p)\rd p/\rd u$. Using $u=g_1(p)$, with $g_1$ defined in (\ref{accuse}), we get $\qf(u|X=1)\propto 1/(1+u^2)$.
Applying the same reasoning to the case $X_{ji}=0$, with $u=g_0(p)$, yields
$\qf(u|X=0)\propto 1/(1+u^2)$.
From the conditional probability we obtain $\qf(u)$ by multiplying with the probability that the event $X=1$ (or $X=0$) occurs,
\be
	\qf(u) = p \qf(u|X=1)=(1-p)\qf(u|X=0)\propto \frac{1}{(1+u^2)^2}.
\label{udist}
\ee
Here we have used the fact that $p=1/(1+u^2)$ for $X=1$ and $p=u^2/(1+u^2)$ for~$X=0$.
Thus the tail of the probability distribution has a $1/u^4$ power law behavior.

Next we argue that the number of contributing terms to $S_j$ (almost $m$ terms for the optimal colluder strategy) is sufficiently large for the CLT to cover 8 sigmas.
For a distribution with vanishing third cumulant and with $\EE[u^4]<\infty$, it is known (see e.g. \cite{Baz2005}) that the region of convergence for the CLT, expressed in sigmas, is given by
\be
	\#{\rm sigmas}=\left(\frac{24\qk_2^2}{\qk_4}\right)^{1/4}N^{1/4},
\label{Nsigmas}
\ee
where $N$ is the number of variables summed, and $\qk_j$ stands for the $j'th$ cumulant.
Our distribution (\ref{udist}) satisfies the requirement $\EE[u^4]<\infty$ because $\qf(u)$ is defined on the finite interval 
$(-\sqrt{\frac{1-t}{t}}, -\sqrt{\frac{t}{1-t}})\cup
 (\sqrt{\frac{t}{1-t}}, \sqrt{\frac{1-t}{t}})$.
We have $\qk_2=1+{\cal O}(\sqrt{t})$ and $\qk_4/\qk_2^2=4/(\pi\sqrt{t})+{\cal O}(1)$.
Substitution into (\ref{Nsigmas}), with $N=m=2\pi^2 c_0^2\ln\qe_1^{-1}$, gives
\be
	\#{\rm sigmas}= (12\pi^3\ln\qe_1^{-1})^{1/4}t^{1/8}\sqrt{c_0}
	\approx 5.2 c_0^{3/8},
\ee
where we have used $\qe_1\approx 10^{-15}$ and $t^{\rm T}=1/(300c_0)$.
Hence, the 8-sigma point of the tail is correctly approximated by a Gaussian already at $c_0\geq 4$.


\section*{Acknowledgments}
We thank Stefan Katzenbeisser and the anonymous reviewers for useful discussions and comments.


\end{document}